\documentclass[a4paper,11pt]{article}
\usepackage[english]{babel}
\usepackage[intlimits]{amsmath}
\usepackage[dvips]{graphicx}

\newcommand{\avg}[1]{\langle #1 \rangle}
\newcommand{\ket}[1]{ | #1 \rangle}
\newcommand{\bra}[1]{ \langle #1 | }

\title{Analysis of trembling hand perfect equilibria in quantum games}
\author{Ireneusz Paku\l a \footnote{\emph{Department of Field Theory and Elementary Particles, Institute of Physics, University of Silesia, ul. Bankowa 14, 40-007 Katowice, Poland}; e-mail: \emph{ipakula@wp.pl}}}

\begin{document}

\maketitle

\begin{abstract}
We analyse Selten's concept of trembling hand perfect equilibria in the context of quantum game theory. We define trembles as mixed quantum strategies by replacing discrete probabilities with probability distribution functions. Explicit examples of analysis are given.\\


\end{abstract}



\section{Introduction}
Multiple applications of game theory and the development of quantum information theory created the combination of both -- the quantum game theory, which extends the classical game theory to the quantum domain. Many potential applications of quantum information processing -- Shor's algorithm, commercially available quantum cryptography and current work on quantum econophysics \cite{PS1} -- open possibilities of future application of quantum game theory to market \cite{PS1,PS2,MPS}, gambling \cite{G1}, social sciences \cite{games_social_norms} (the cooperation of players as the entanglement of their states), decision science \cite{busemeyer} and the information theory itself, by putting information processing problems into the realm of quantum decision theory \cite{meyer}.

Implementation of a quantum game or a quantum algorithm involves problems due to technical limitations as well as the quantum theory itself and we can hardly hope for perfectness in quantum state preparation and implementation of quantum gates (strategies) -- every real apparatus generates noise \cite{NFJ,toor}. In the case of a game there arises a question of stability of equilibria -- do the 'solutions' of a game survive when uncertainties in the application of strategies (or quantum gates) appear? Several forms of stability of equilibria are known in the game theory, depending on the definition of perturbation which the game is undergoing, for example Selten's trembling hand perfectness \cite{selten}, Myerson's properness \cite{myerson} or Mertens stability \cite{mertens}. In this paper we analyse trembling hand perfectness of quantum equilibria with respect to trembles coming from uncertainty of strategies performed by players. This problem is one of the keystones of implementation of quantum games.

\section{Trembling hand perfectness -- classical case}

The notion of the trembling hand perfect equilibrium was put forward by Selten as a refinement of Nash equilibrium (NE). An equilibrium is \emph{trembling hand perfect (THP)} if there exists a sequence of perturbed equilibria converging to it. Perturbations are given by small probabilities of playing non-equilibrium strategies -- a player plays a completely mixed strategy instead of playing a pure equilibrium one (we may say hands of the players are trembling therefore they make mistakes in the choice of their strategies during the game). To be precise, a strategy profile $\sigma$ is a trembling hand perfect equilibrium if there exists a sequence of totally mixed strategy profiles $\sigma^n \to \sigma$ such that
\begin{equation}
\bigwedge_i \bigwedge_{s_i\in S_i} \$_i(\sigma_i,\sigma_{-1}^n)\geq \$_i(s_i,\sigma_{-1}^n),
\end{equation}
where $\$_i$ is the payoff function for the $i$th player, depending on strategies $\sigma_i$ and $\sigma_{-i}$ (which represents strategies from the strategy profile $\sigma$ for players other than $i$) and $s_i$ is a pure strategy from the strategy space $S_i$  \cite{gameth_ft}.

As an example of trembling hand perfect and imperfect equilibria we use the game (later on referred to by EG) with the payoff bimatrix:

\begin{center}
\begin{tabular}{cc}
& B \\
A & 
\begin{tabular}{c|c|c}
& C & D \\
\hline
C & (1,1) & (2,0) \\
\hline
D & (0,2) & (2,2)
\end{tabular}
\end{tabular}
\end{center}

The players -- Alice (A) and Bob (B) -- both can use strategy $C$ or $D$. In the payoff bimatrix above left numbers represent payoffs for Alice, right ones -- for Bob. The game is symmetric and has two Nash equilibria -- $(C,C)$ and $(D,D)$. Let us calculate expected payoffs for players, when the other one plays his strategy perturbed by 'trembles'.

When we analyse the first equilibrium, Alice plays a mixed strategy $\sigma_A=(p_C,p_D)=(1-\epsilon,\epsilon)$ (strategy $C$ perturbed by trembles -- she plays $D$ with small probability $\epsilon$) and Bob's expected payoffs from playing $C$ and $D$ are given by:
\begin{eqnarray}
C: \$_B(\sigma_A,C)&=&1(1-\epsilon)+2\epsilon=1+\epsilon, \\
D: \$_B(\sigma_A,B)&=&0(1-\epsilon)+2\epsilon=2\epsilon.
\end{eqnarray}
As we can see, for small values of $\epsilon$ it is best for Bob to play $C$. Because the game is symmetric, Alice encounters the same situation when Bob's hand playing strategy $C$ is trembling -- and $(C,C)$ is a trembling hand perfect equilibrium (THP). If we perform similar calculations for $(D,D)$ we find that when a player plays $D$ 'contaminated' with $C$, it is best for his opponent to play $C$ -- the equilibrium $(D,D)$ is not trembling hand perfect (THiP).

\section{Quantum games -- mixed and generalised continuously mixed strategies -- trembles}
We use mixed states to introduce mixed strategies. Usually the problem is formulated in the following way \cite{eisert_thesis}: the Alice's and Bob's strategies
\begin{eqnarray}
A_i&=&A(\theta^A_i,\alpha^A_i,\beta^A_i)=A(\Omega^A_i)\\
B_j&=&B(\theta^B_j,\alpha^B_j,\beta^B_j)=B(\Omega^B_j),
\end{eqnarray}
together with the payoff operator (see Appendix A) give us the expected payoff functions:
\begin{eqnarray}
\avg{\$_A(s^A_i,s^B_j)}&=&Tr(P_A \rho_f(A_i,B_j))\\
\avg{\bar{\$}_A}&=&\sum_{i,j}p^A_i p^B_j \avg{\$_A(s^A_i,s^B_j)}
\end{eqnarray}
where $\Omega^{A(B)}_i$ is the set of parameters of the unitary strategy Alice (Bob) uses and the bar means the strategies are mixed. As we need completely mixed strategies to analyse trembling hand perfectness of equilibria, we use probability distribution functions (PDFs) $f_{A(B)}(\Omega)$ instead of discrete probabilities. We integrate the payoff functions over all strategy spaces with measure $\mu$:
\begin{multline}
\avg{\bar{\$}_A}=\int_{SU(2)\times SU(2)}f_A(\Omega^A,\Omega^A_0)f_B(\Omega^B,\Omega^B_0)\times \\ \times\avg{\$_A(A(\Omega^A),B(\Omega^B))}\mu(\Omega^A)\mu(\Omega^B) \label{eq:int_cont_mix}
\end{multline}
The whole strategy space is the space of quantum operations -- trace preserving positive maps. However we restrict ourselves to (special, as overall phases do not change values of payoff functions) unitary operations. Furthermore, due to symmetries of the payoff functions, we use $U\in SU(2)$ matrices of the form given below, with $\theta\in [-\pi,\pi],\alpha\in [0,2\pi],\beta\in [0,2\pi]$. As our resulting strategy space is the torus $S^1\times S^1\times S^1$, we use von Mises distributions ($S^1$ version of vMF PDF -- see Appendix B) in each parameter space.
\begin{eqnarray}
U(\theta,\alpha,\beta)&=&\left[\begin{array}[c]{ll}
e^{i\alpha/2}\cos{\theta/2} & e^{i\beta/2}\sin{\theta/2} \\
-e^{-i\beta/2}\sin{\theta/2} & e^{-i\alpha/2}\cos{\theta/2}
\end{array}\right] \label{eq:su2matrix}\\
c_2&=&\frac{1}{2\pi I_0(\kappa)} \\
f(\Omega,\Omega_0)&=&c_2^3\exp[\kappa(\cos(\theta-\theta_0)+\cos(\alpha-\alpha_0)+\cos(\beta-\beta_0))],
\end{eqnarray}
where $c_2$ is the $S^1$ normalisation factor.
The integral (\ref{eq:int_cont_mix}) gives us the expected payoff in the case of two initially pure strategies smeared over the whole $SU(2)$. This method of perturbing can be easily extended to the case of classical discretely mixed strategies (the integral parameters are suppressed due to their obviousness):
\begin{equation}
\avg{\bar{\$}_A}=\sum_{i,j}p^A_i p^B_j \int_{SU(2)\times SU(2)}f^A_i f^B_j\avg{\$_A(A,B)}\mu_A\mu_B.
\end{equation}

For $n\times n$ games, we need higher dimensional strategy spaces, namely $SU(n)$, so instead of using qubits, we need to speak in the language of qunits. For N players the expected payoff function appears of the form:
\begin{multline}
\avg{\bar{\$}_A}=\idotsint_{SU(n)\times\ldots\times SU(n)}f_A(\Omega^A,\Omega^A_0)f_B(\Omega^B,\Omega^B_0)\ldots f_N(\Omega^N,\Omega^N_0)\times\\
\times\avg{\$_A(A(\Omega^A),B(\Omega^B),\ldots,N(\Omega^N))}\mu(\Omega^A)\mu(\Omega^B)\ldots\mu(\Omega^N),
\end{multline}
where, in the case of mixed strategies needed to analyse trembling hand perfectness of equilibria, probability distribution functions would be $SU(N)$ analogues of Gauss or rather von Mises-Fisher distributions.

\section{Analysis}

We perturb pure strategies by smearing them over the whole strategy space using certain probability distribution functions, which go smoothly to the pure case (Dirac's delta distribution) in the limit. Changes of the parameter describing our perturbations do not change the qualitative properties of the shape of the payoff functions (except for the Stag Hunt game, where the change is shown in detail), thus only one example of perturbed function in each case is presented below, where the properties of the shape are clearly visible.

\subsection{The Prisoners' Dilemma}

\begin{figure}[x]
\centerline{
\includegraphics[width=3.0in,height=2.5in]{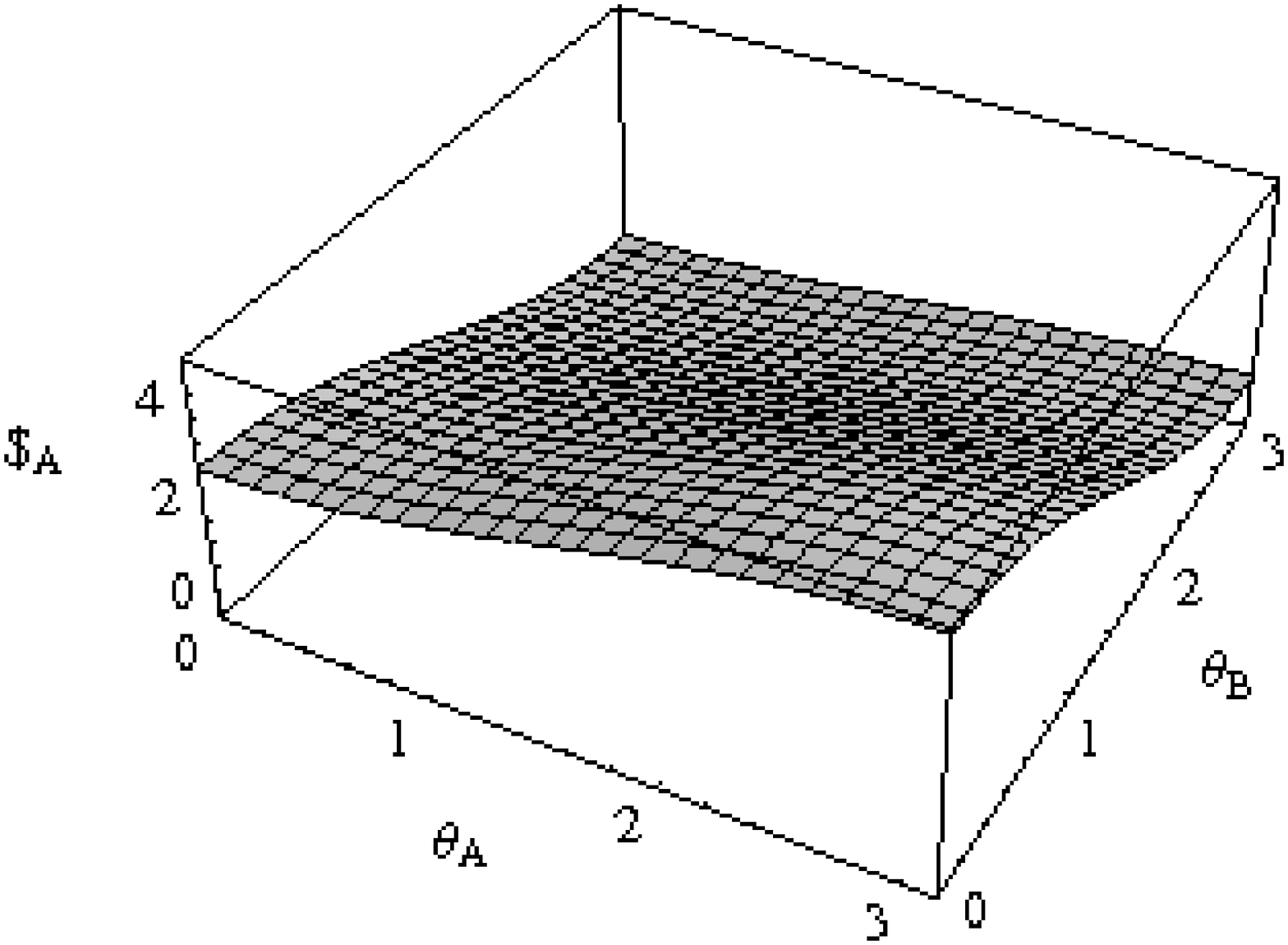}
\includegraphics[width=3.0in,height=2.5in]{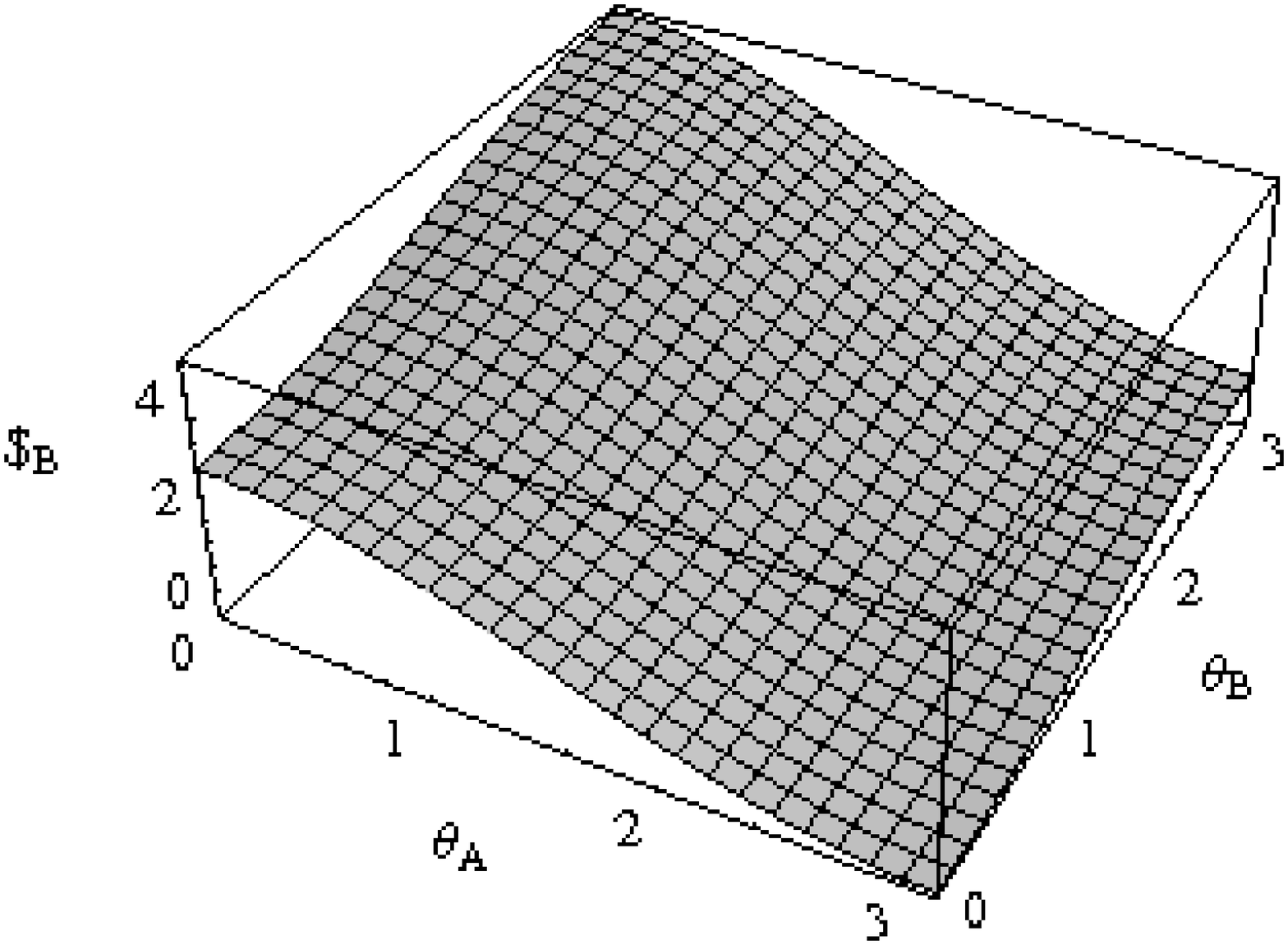}
}
\caption{PD: Payoff for Alice and Bob playing pure strategies against pure opponent strategies, classical case (1 parameter -- mixed strategies)}\label{fig:PD_klas_miesz}
\end{figure}

The Prisoner's Dilemma (PD) is a commonly used game with the payoff bimatrix given by:
\begin{center}
\begin{tabular}{cc}
& B \\
A & 
\begin{tabular}{c|c|c}
& C & D \\
\hline
C & (3,3) & (0,5) \\
\hline
D & (5,0) & (1,1)
\end{tabular}
\end{tabular}
\end{center}
When we analyse this classical game we acquire the following mixed strategy payoffs (one parameter strategy space) -- fig. \ref{fig:PD_klas_miesz}. 

\begin{figure}[x]
\centerline{
\includegraphics[width=3.0in,height=2.5in]{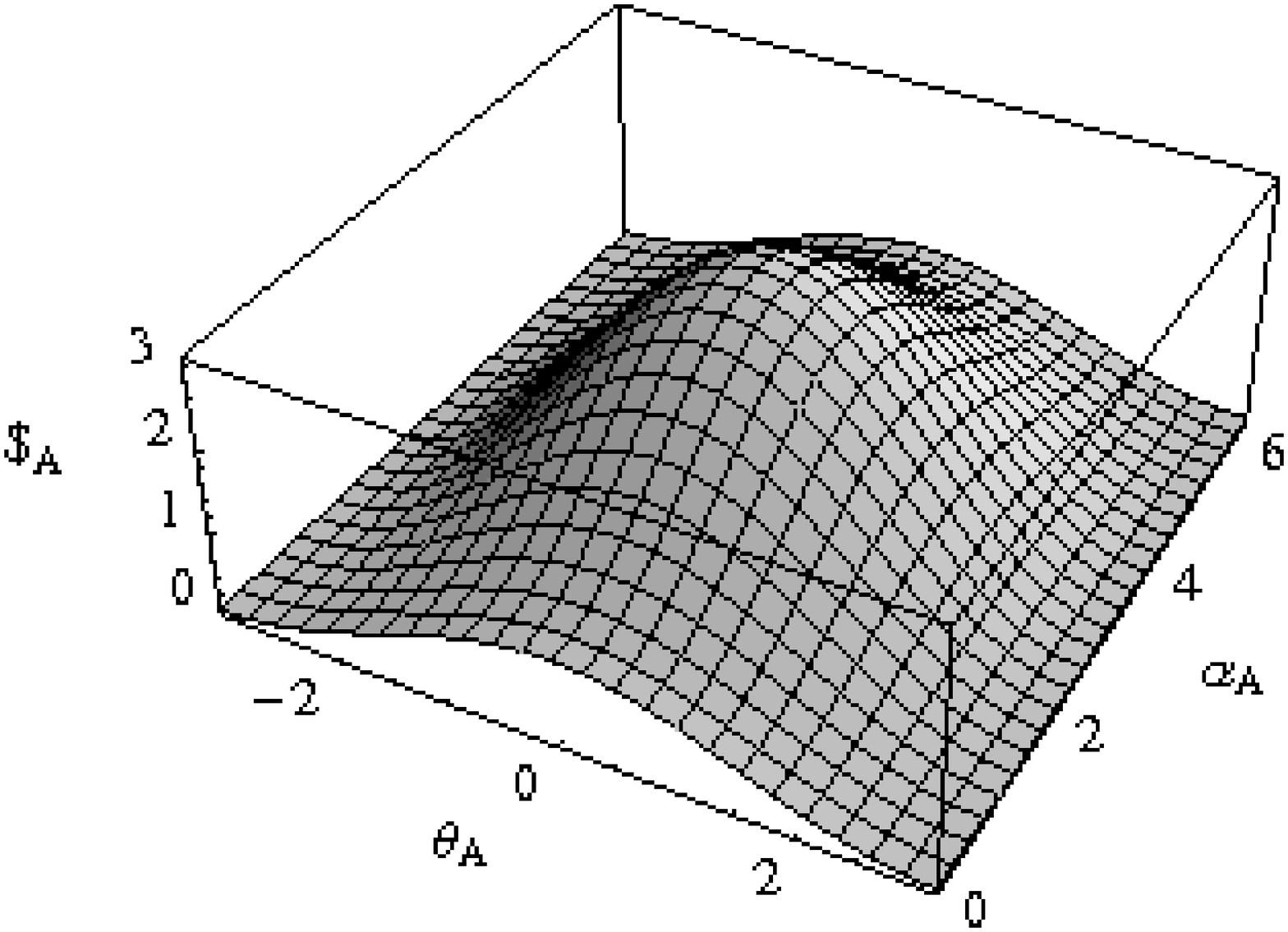}
\includegraphics[width=3.0in,height=2.5in]{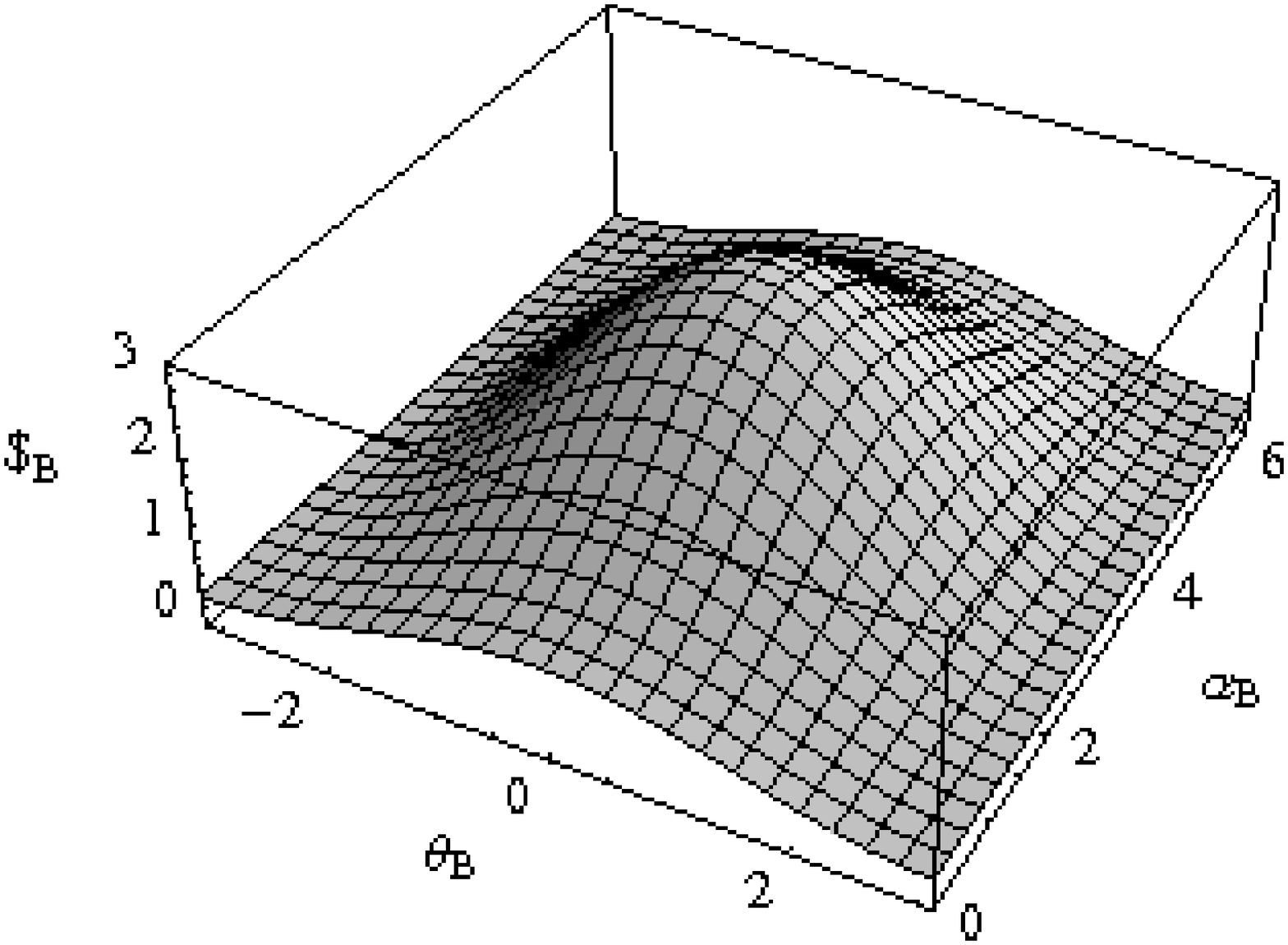}
}
\caption{PD, 2 parameters. Left: Payoff for Alice playing pure strategy against Bob playing pure Q. Right: Payoff for Bob's pure strategy against Alice's trembling Q, $\kappa=5$}\label{fig:PD_2par_BpQAp_BpAtQ}
\end{figure}

In the case of two parameters we restrict the strategy space by taking $\beta=0$ in formula (\ref{eq:su2matrix}) $SU(2)$ matrix thus getting the torus $S^1\times S^1$. Then we use vM PDF for $S^1$ parameter spaces to introduce trembles. We observe stability of Nash equilibrium given by the strategy profile $(Q,Q)$, where $Q$ is a strategy of the form:
\begin{equation}
Q=U(0,\pi,0)=\left[\begin{array}[c]{ll}
i & 0 \\
0 & -i
\end{array}\right]. \label{eq:Q}
\end{equation}
For the case of pure Bob's strategy $B=Q$ and Alice's strategy $A$ unperturbed, the equilibrium point is clearly seen as a maximum of Alice's payoff function (fig. \ref{fig:PD_2par_BpQAp_BpAtQ} -- left).
For all cases of $\kappa$ we investigate, the payoff functions are preserving their shape with maximum representing the Nash equilibrium (fig. \ref{fig:PD_2par_BpQAp_BpAtQ} -- right). This behaviour is present when both Alice's and Bob's strategies are disturbed by trembles as well.

\begin{figure}[x]
\centerline{
\includegraphics[width=3.0in,height=2.5in]{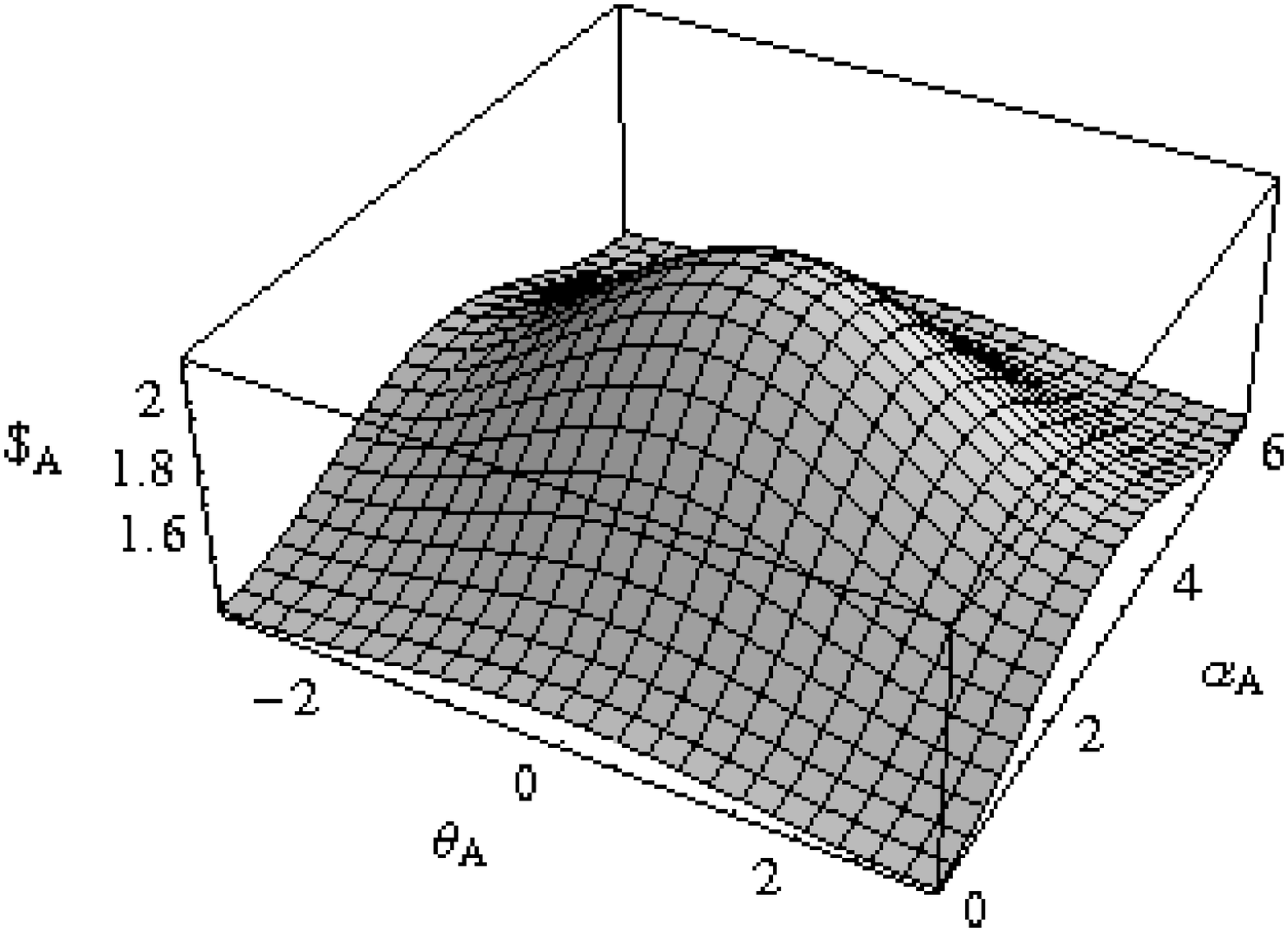}
\includegraphics[width=3.0in,height=2.5in]{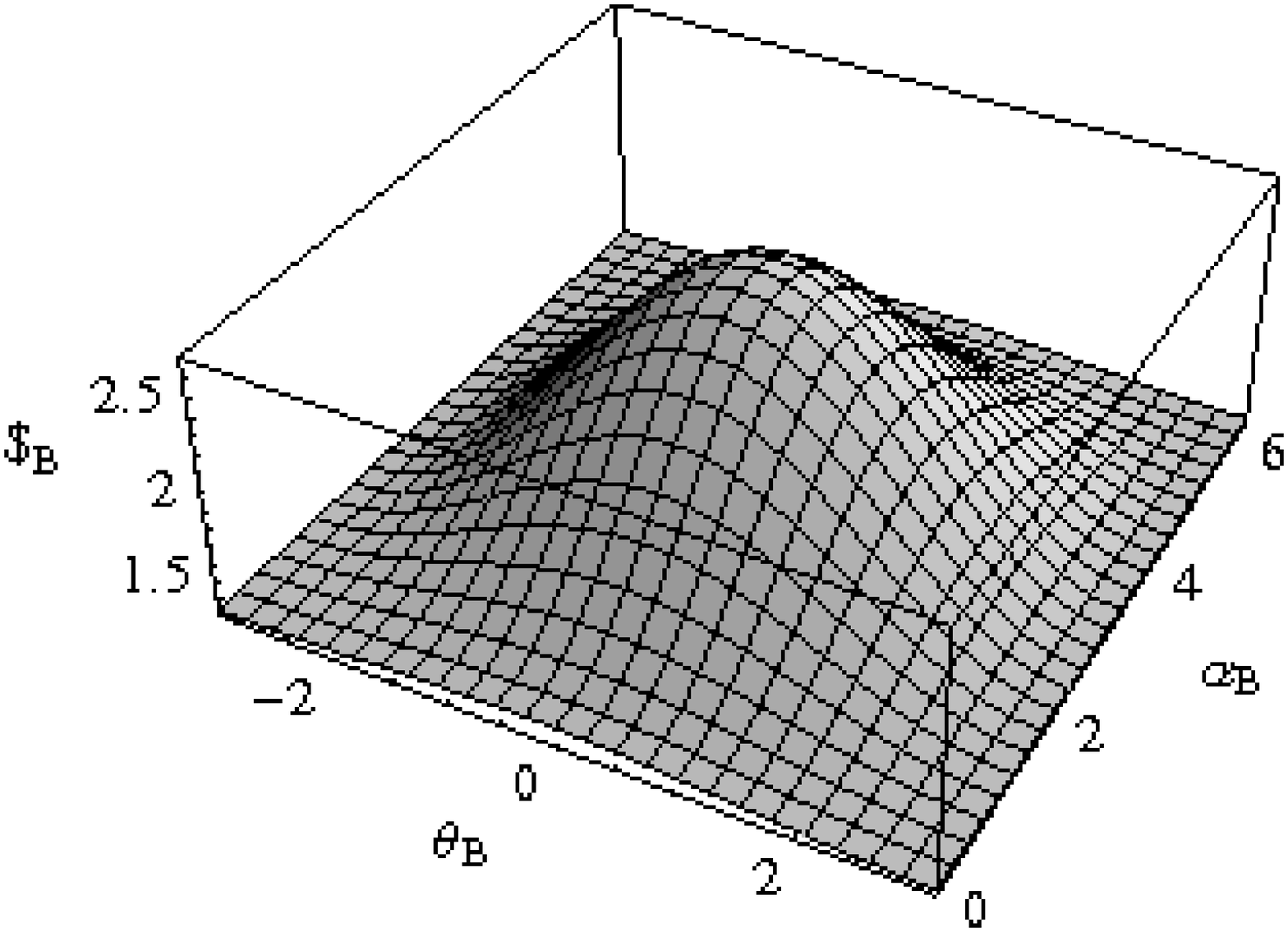}
}
\caption{PD, 3 parameters: Payoff for Alice (left) and Bob (right) when Alice plays trembling strategies against pure Bob's Q, $\kappa=1$}\label{fig:PD_3par_PA_PB_BpAtQ}
\end{figure}

Next we leave pure Bob's strategy within the two parameter strategy space while Alice's trembling hand smears her strategy over the full three parameters strategy space.
In spite of trembling Alice's strategy, payoff functions preserve their qualitative properties (fig. \ref{fig:PD_3par_PA_PB_BpAtQ}).

\subsection{An example of a $2\times 2$ game -- quantum case}

\begin{figure}[x]
\centerline{
\includegraphics[width=3.0in,height=2.5in]{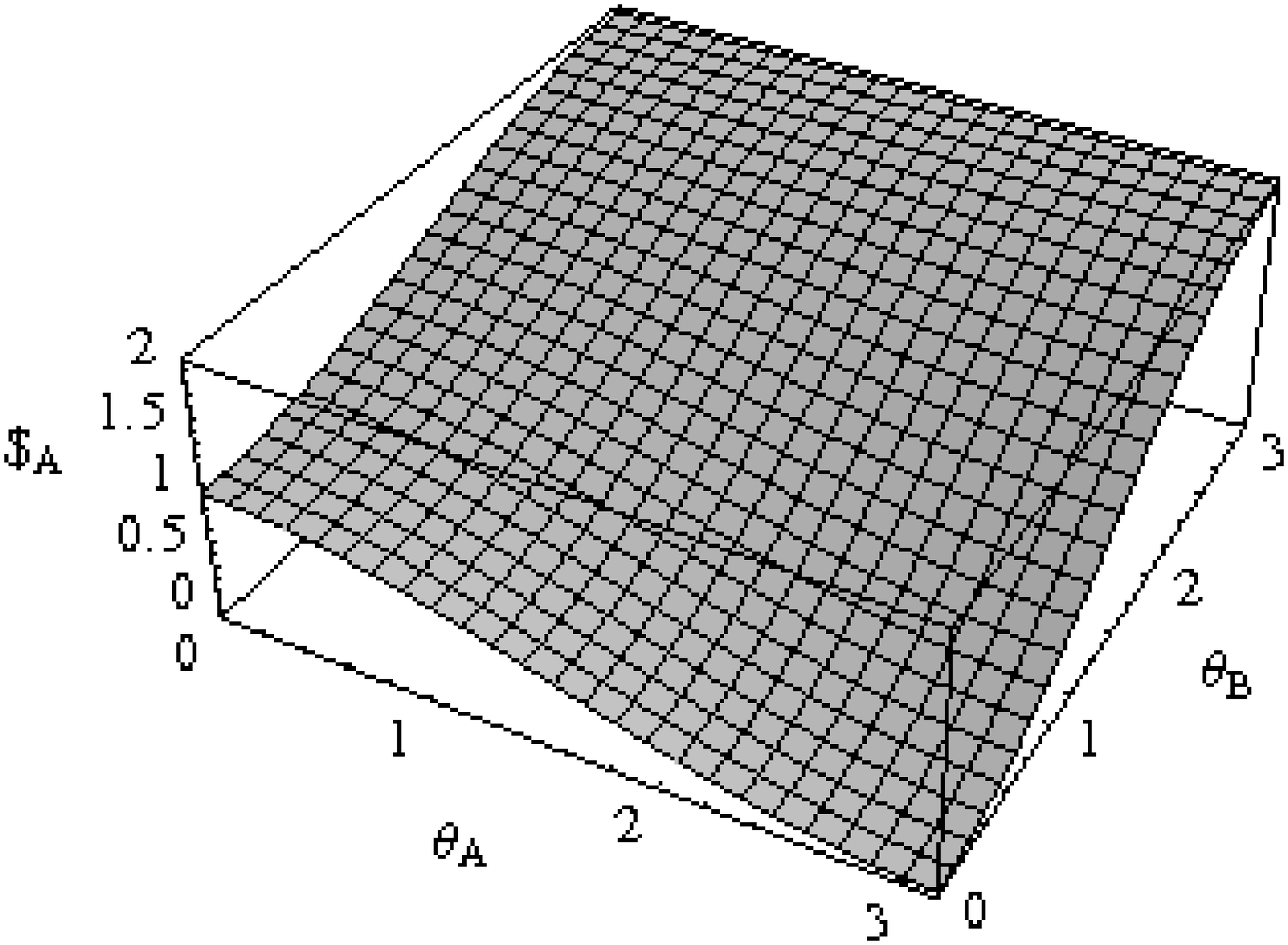}
\includegraphics[width=3.0in,height=2.5in]{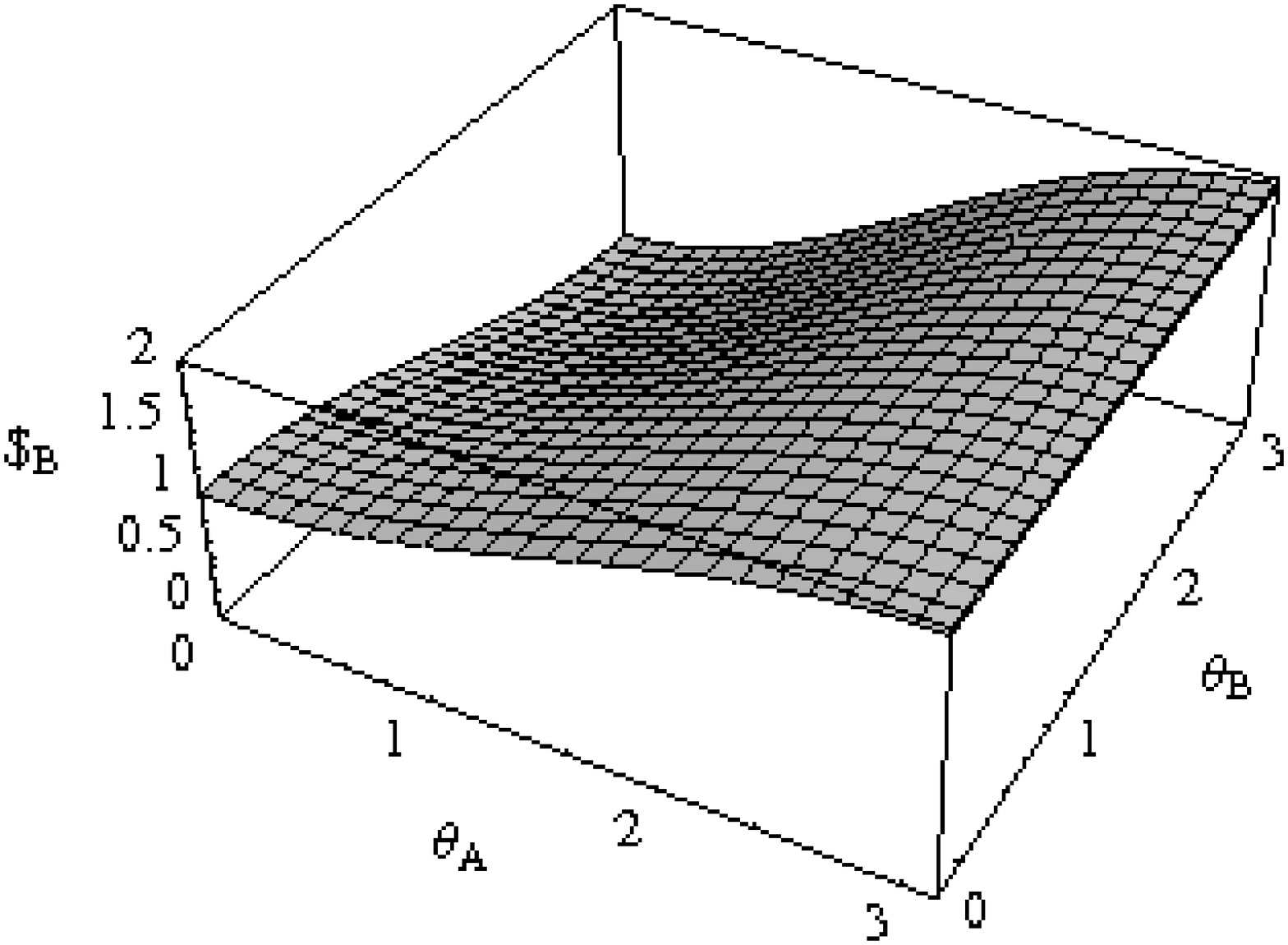}
}
\caption{EG: Payoff for Alice and bob playing pure strategies against pure opponent strategies, classical case (1 parameter -- mixed strategies)}\label{fig:EG_klas_miesz}
\end{figure}

\begin{figure}[x]
\centerline{
\includegraphics[width=3.0in,height=2.5in]{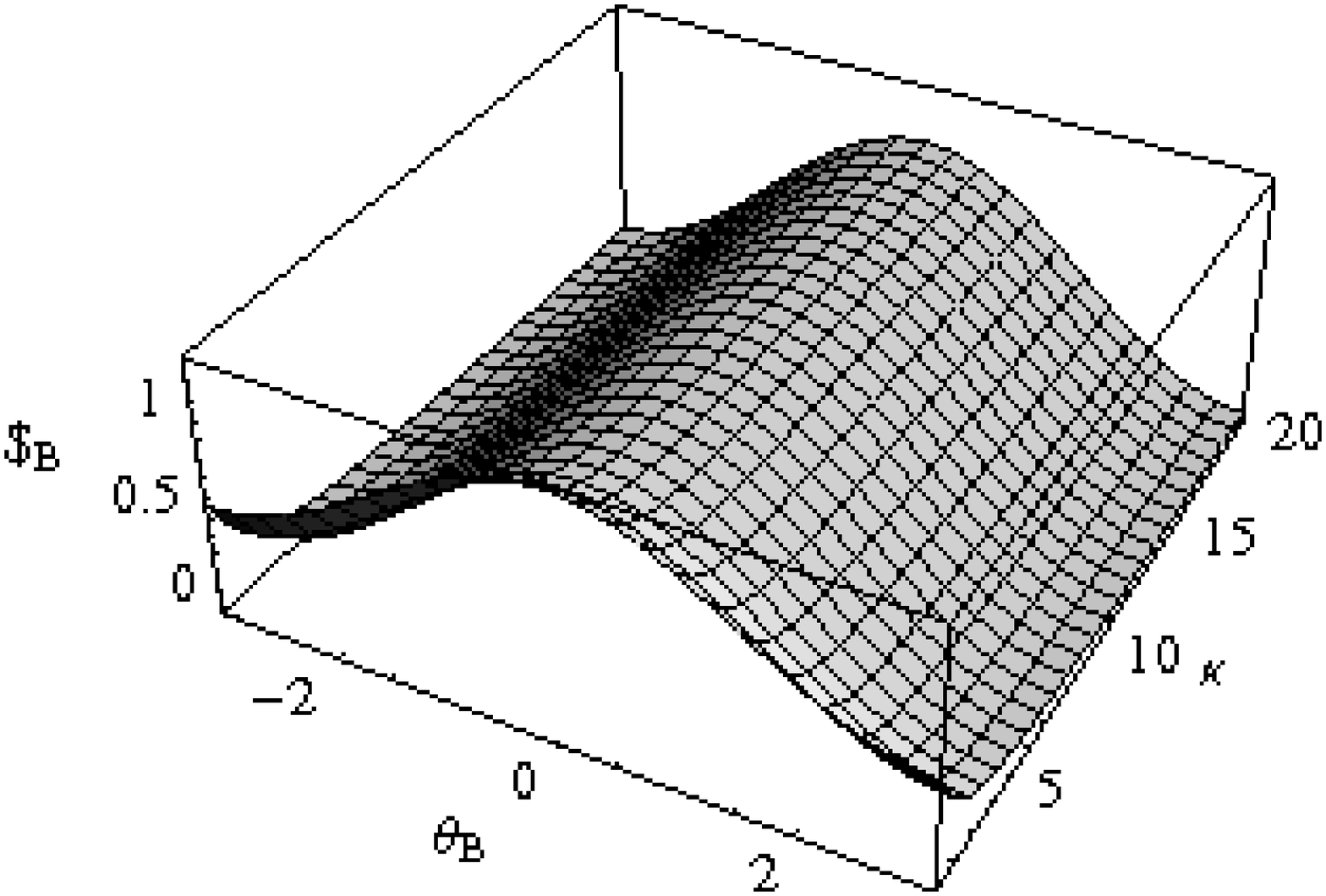}
\includegraphics[width=3.0in,height=2.5in]{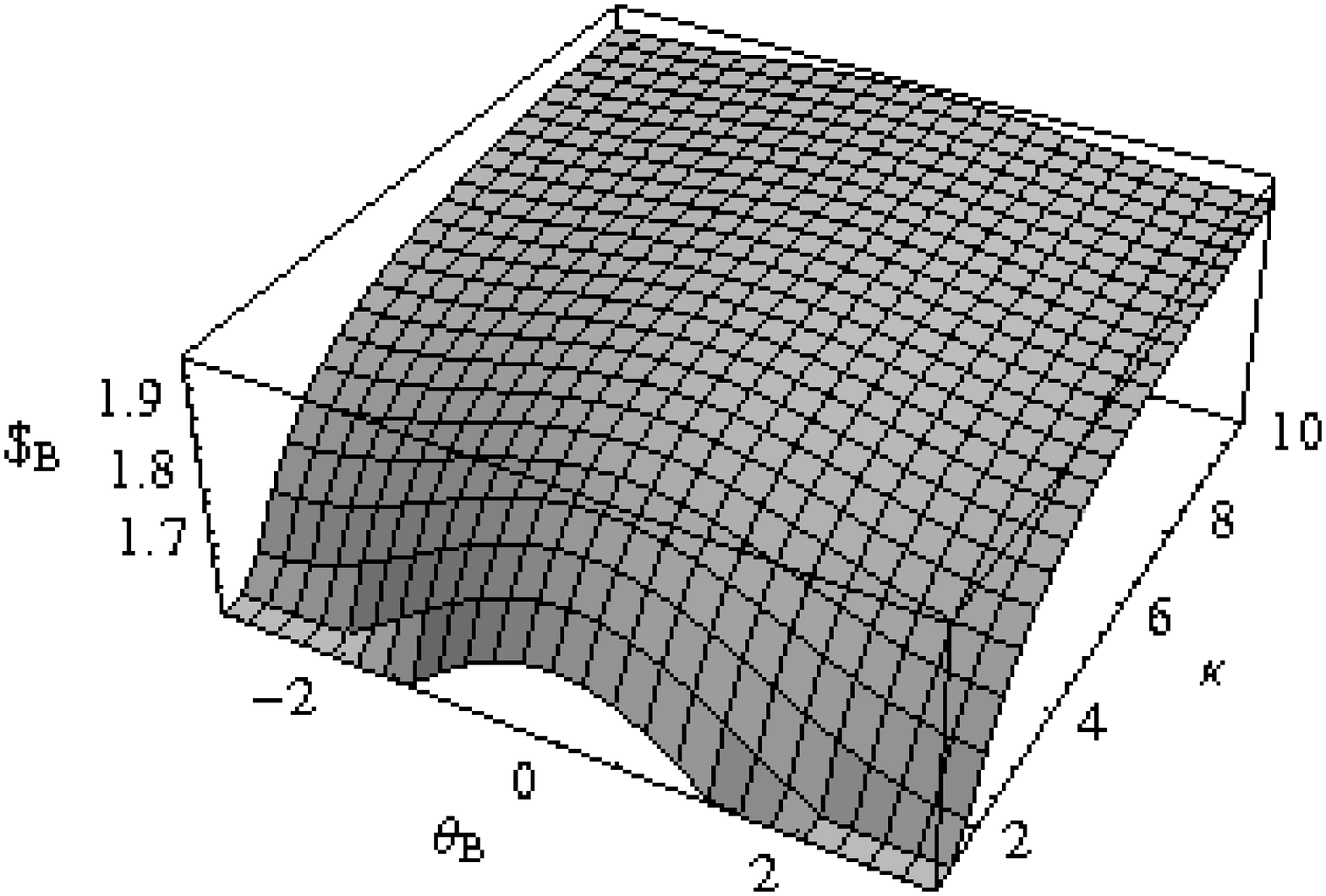}
}
\caption{EG. Left: Perfection of classical equilibrium: Alice plays trembling C, Bob plays pure strategy. Right: Imperfection of classical equilibrium: Alice plays trembling D, Bob plays pure strategy.}\label{fig:EG_THP_THiP}
\end{figure}

As we previously mentioned EG possesses two classical equilibria, one of which is THiP. Mixed strategy payoffs for this game are given by figures (\ref{fig:EG_klas_miesz}). Figures (\ref{fig:EG_THP_THiP}) present $\$_B$ as a function of Bob's strategy and the concentration parameter. The persistent maximum for $(C,C)$ ($\theta=0$) is clearly seen, as well as the imperfectness of $(D,D)$ ($\theta=\pi$) -- it is preferred for Bob to change his strategy $D$ for $C$ for all values of $\kappa$ under investigation.

\begin{figure}[x]
\centerline{
\includegraphics[width=3.0in,height=2.5in]{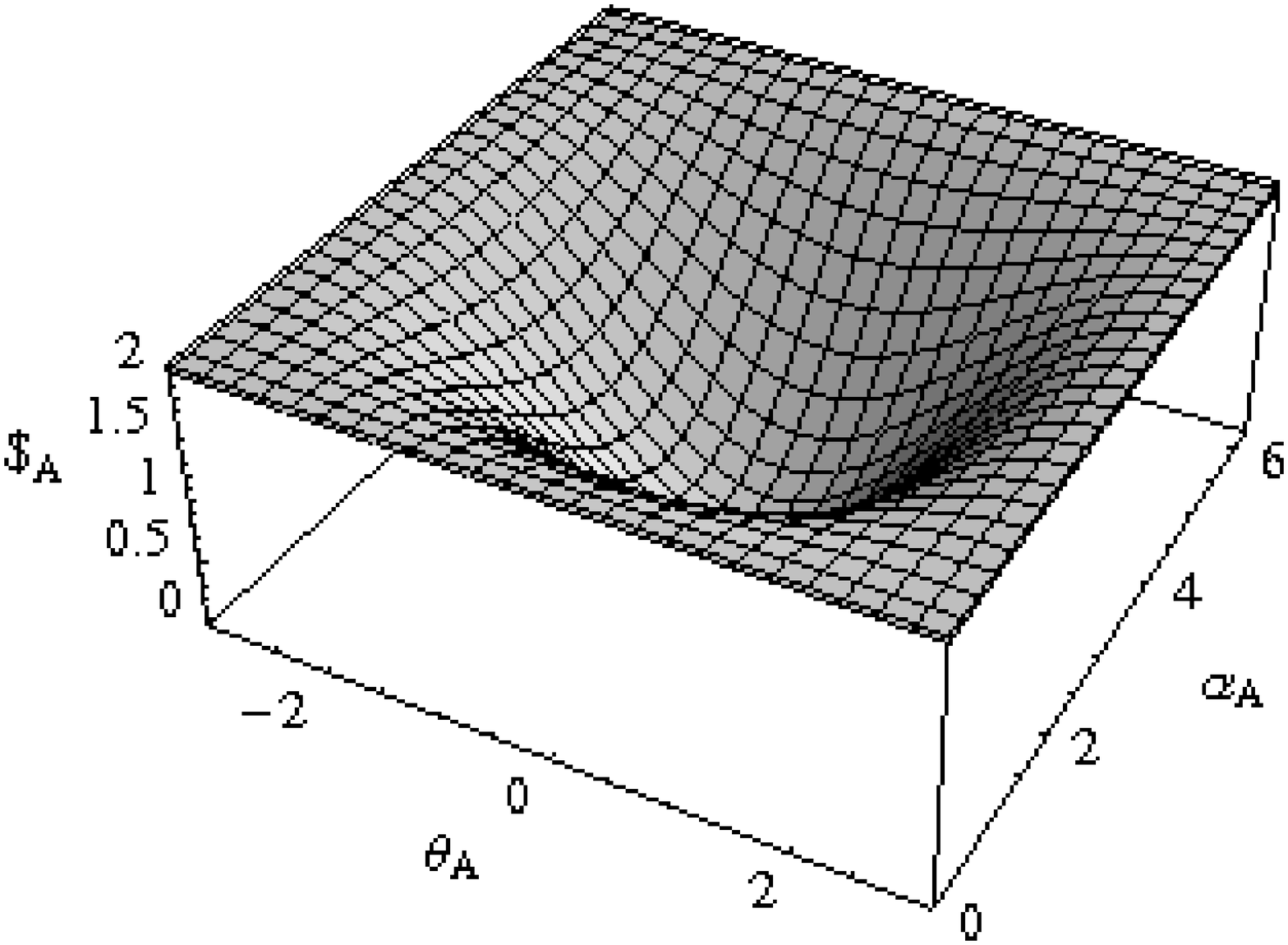}
\includegraphics[width=3.0in,height=2.5in]{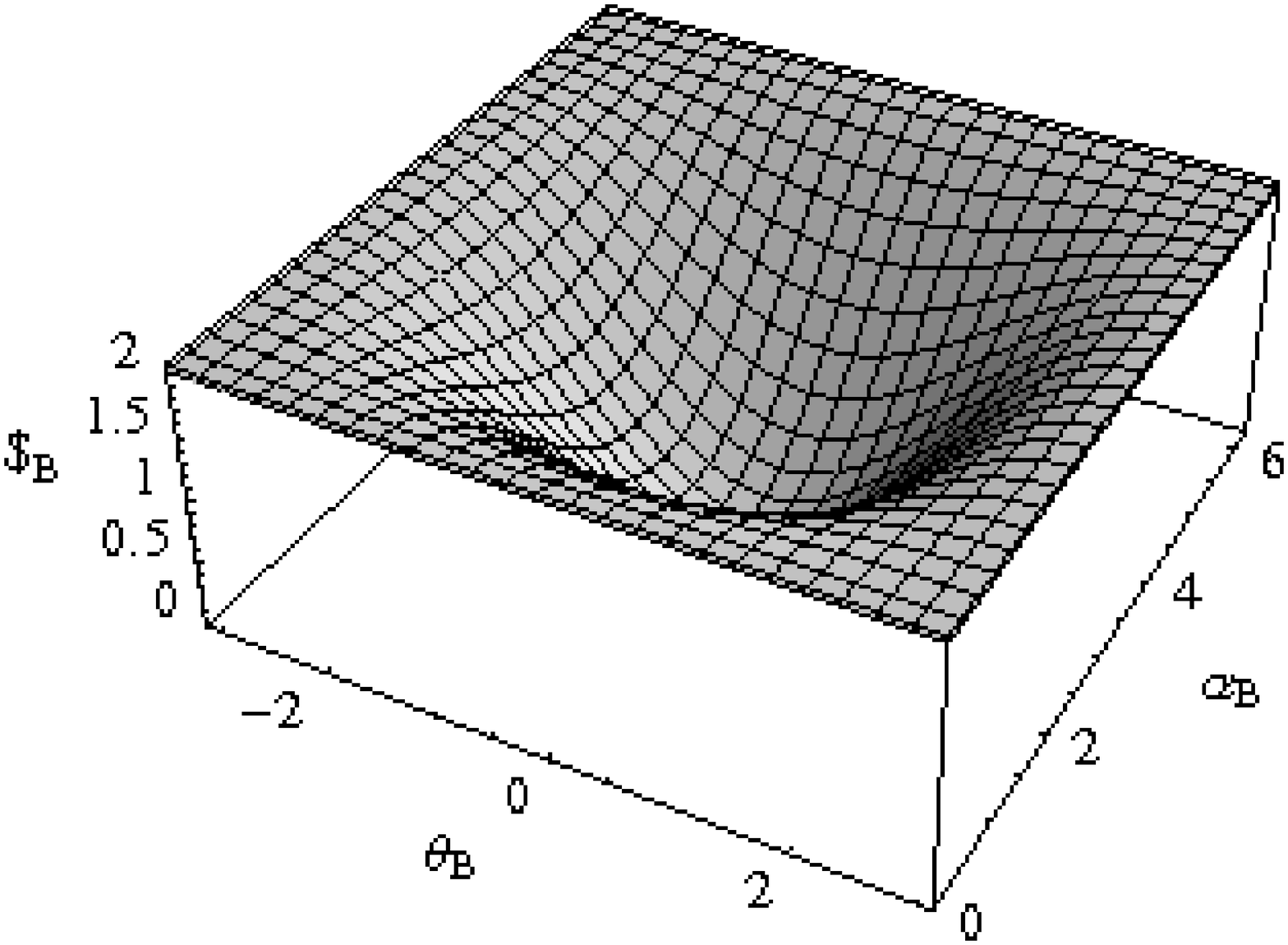}
}
\caption{EG, 2 parameters: Payoff for Alice playing pure strategies against pure Bob's D; payoff for Bob playing pure strategies against pure Alice's D}\label{fig:EG_2par_BpDApD}
\end{figure}

\begin{figure}[x]
\centerline{
\includegraphics[width=3.0in,height=2.5in]{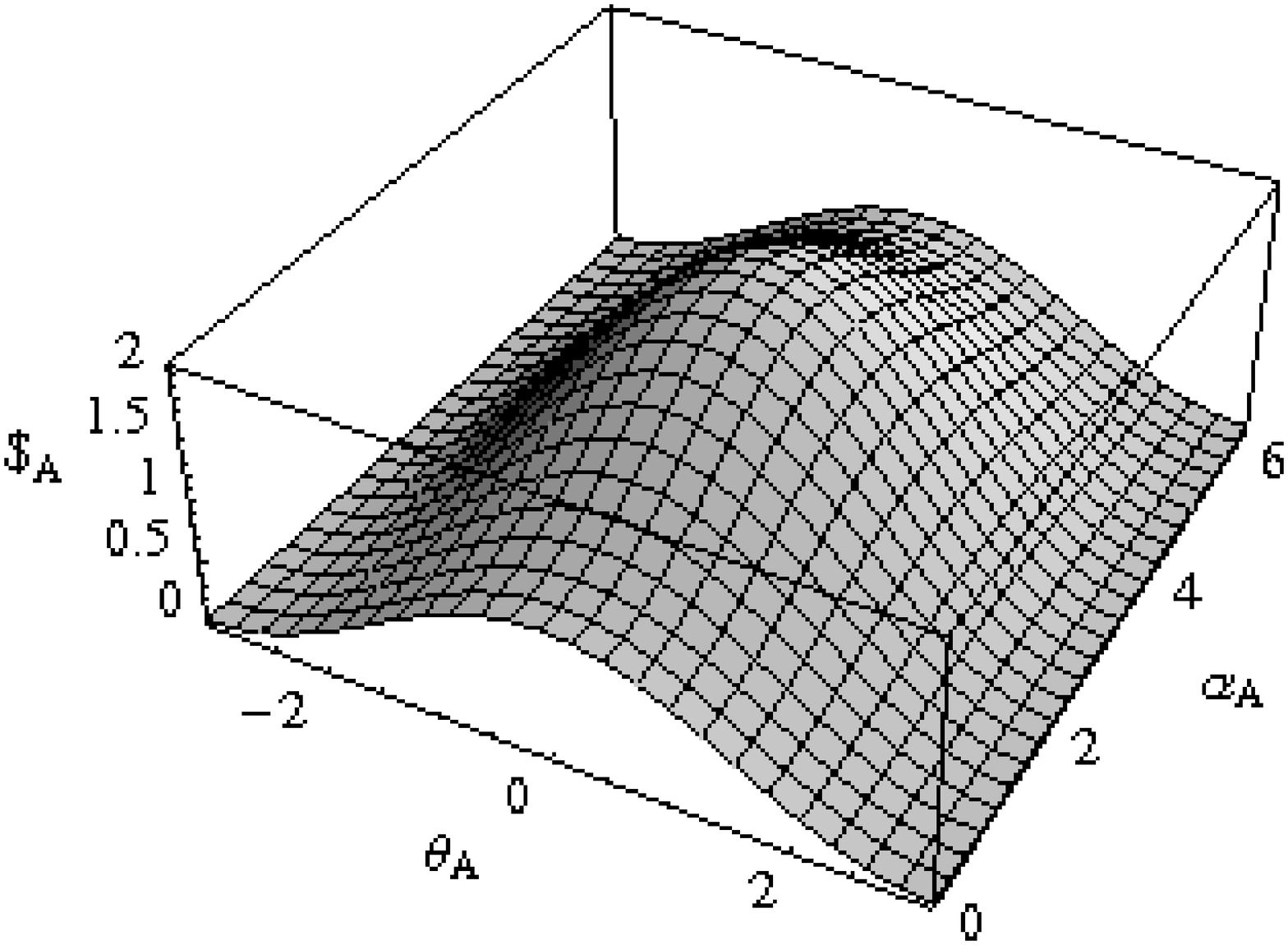}
\includegraphics[width=3.0in,height=2.5in]{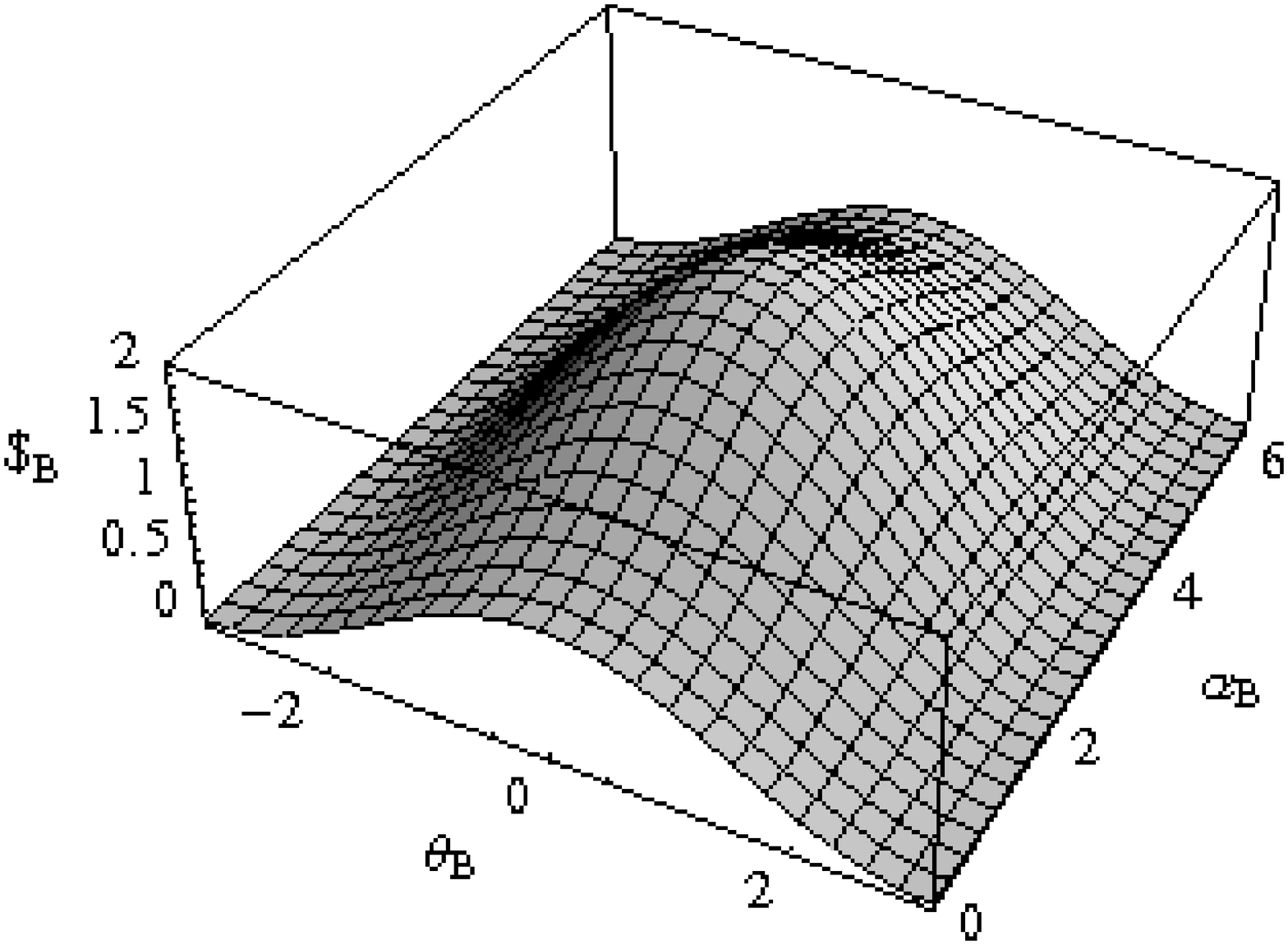}
}
\caption{EG, 2 parameters: Payoff for Alice playing pure strategies against pure Bob's C; payoff for Bob playing pure strategies against pure Alice's C}\label{fig:EG_2par_BpCApC}
\end{figure}

\begin{figure}[x]
\centerline{
\includegraphics[width=3.0in,height=2.5in]{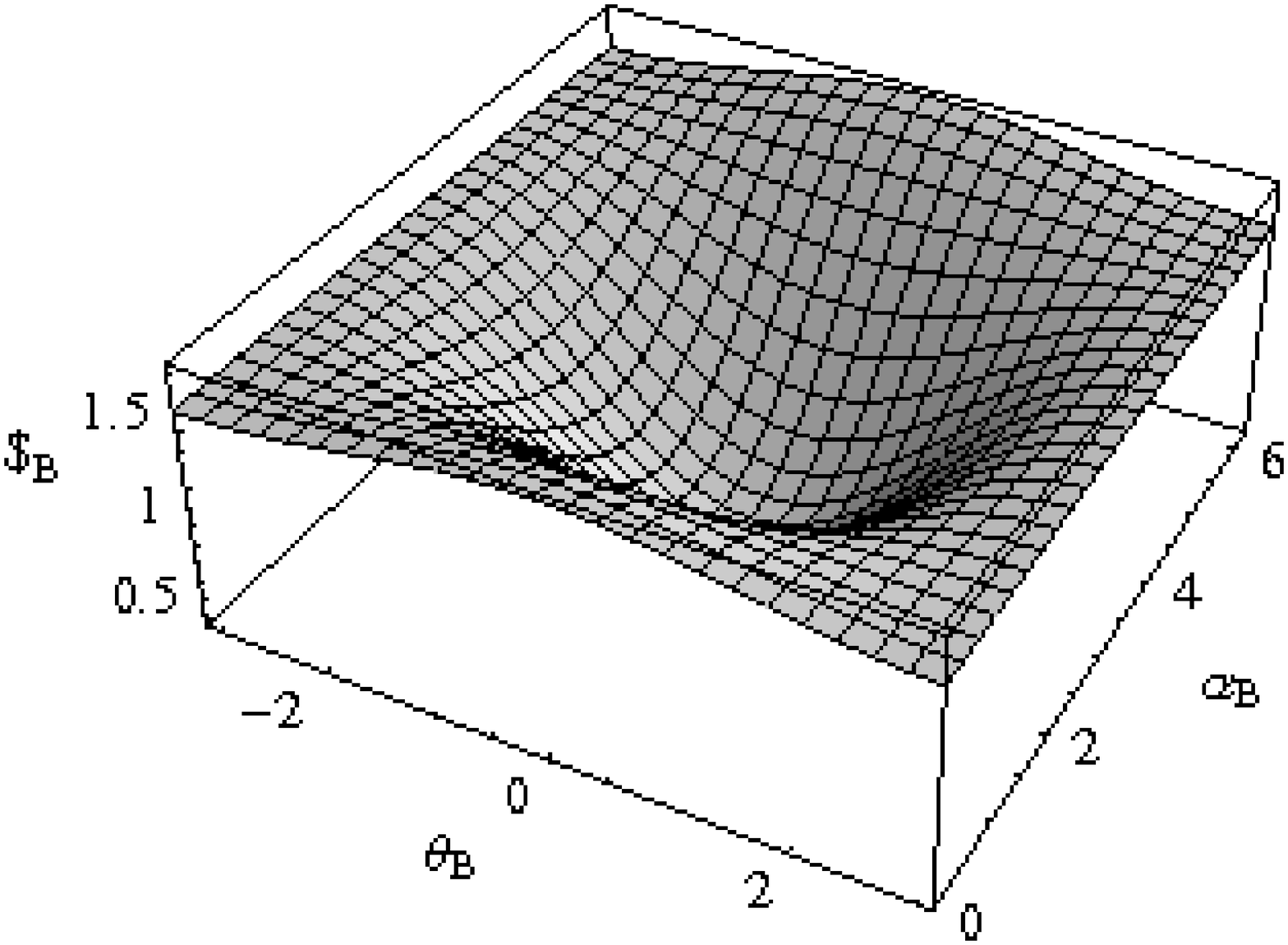}
}
\caption{EG, 2 parameters: Payoff for Bob playing pure strategies against trembling Alice's D, $\kappa=1$}\label{fig:EG_2par_BpAtD}
\end{figure}

After extending our strategy space to two parameters one classical equilibrium remains -- $(D,D)$ -- as a weak maximum of payoff function (fig. \ref{fig:EG_2par_BpDApD}). The strategy profile $(C,C)$ ceases to be an equilibrium in the quantum case (fig. \ref{fig:EG_2par_BpCApC}). However, when the opponent's strategy is trembling in a two parameter space the strategy $D$ is not an equilibrium anymore, for it is better to play $C$ (fig. \ref{fig:EG_2par_BpAtD}).

\begin{figure}[x]
\centerline{
\includegraphics[width=3.0in,height=2.5in]{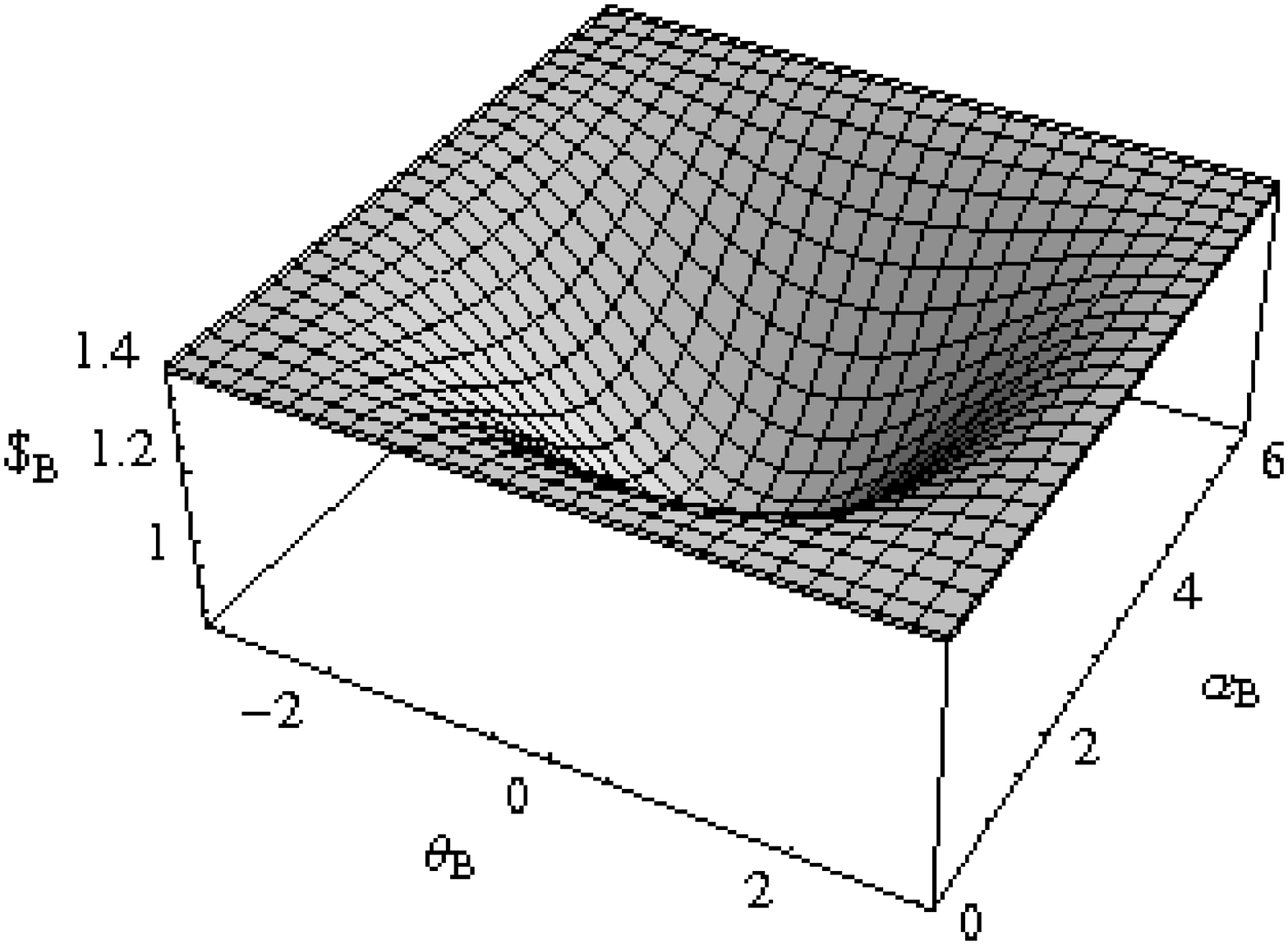}
}
\caption{EG, 3 parameters: Payoff for Bob playing pure strategies against trembling Alice's D, $\kappa=1$}\label{fig:EG_3par_BpAtD}
\end{figure}

Surprisingly, when we allow Alice's hand to tremble in the three parameter space the strategy profile $(D,D)$ persists to be an equilibrium -- trembles just flatten the payoff function without changing its qualitative properties (fig. \ref{fig:EG_3par_BpAtD}).


\subsection{The Stag Hunt}

\begin{figure}[x]
\centerline{
\includegraphics[width=3.0in,height=2.5in]{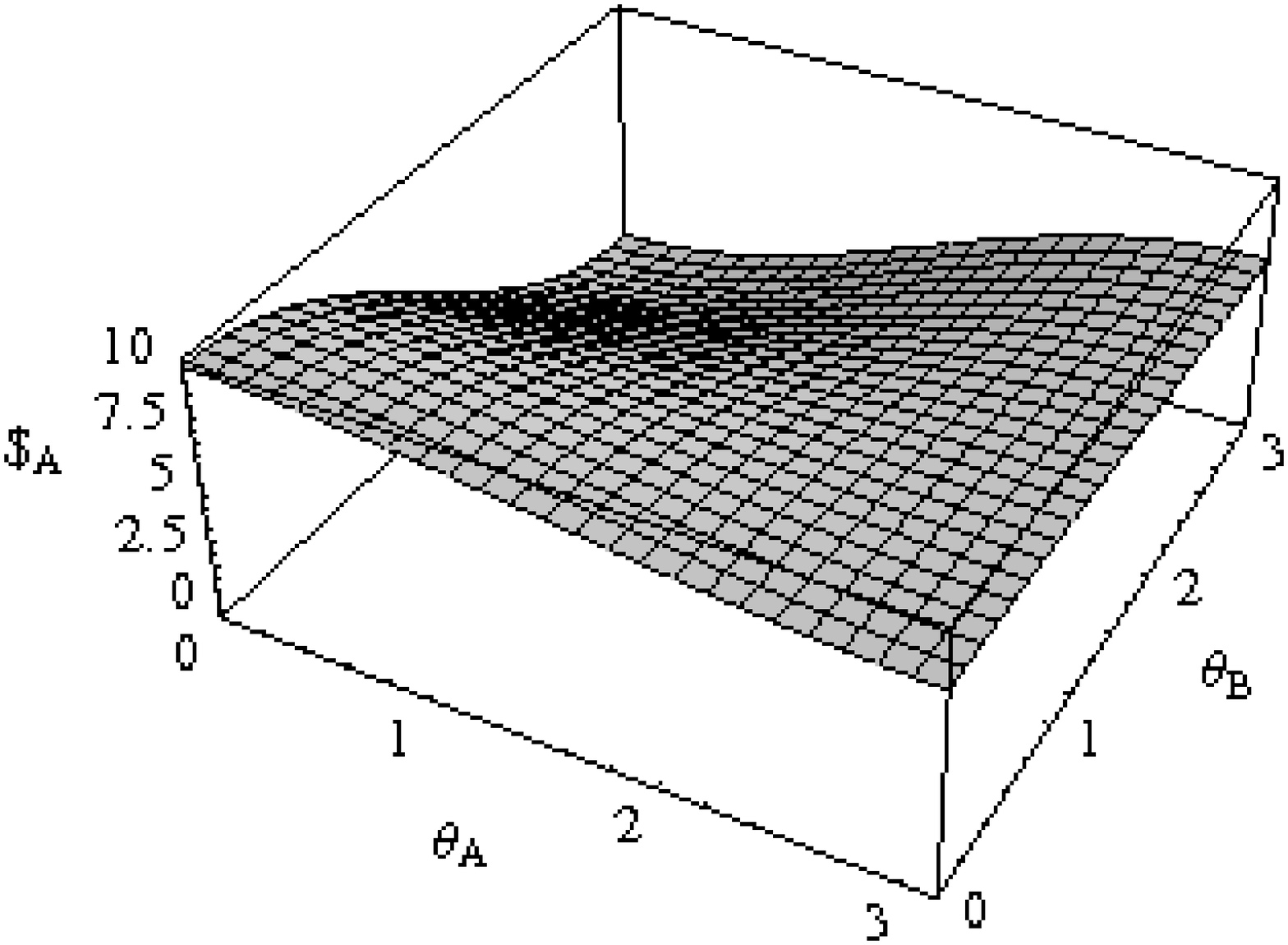}
\includegraphics[width=3.0in,height=2.5in]{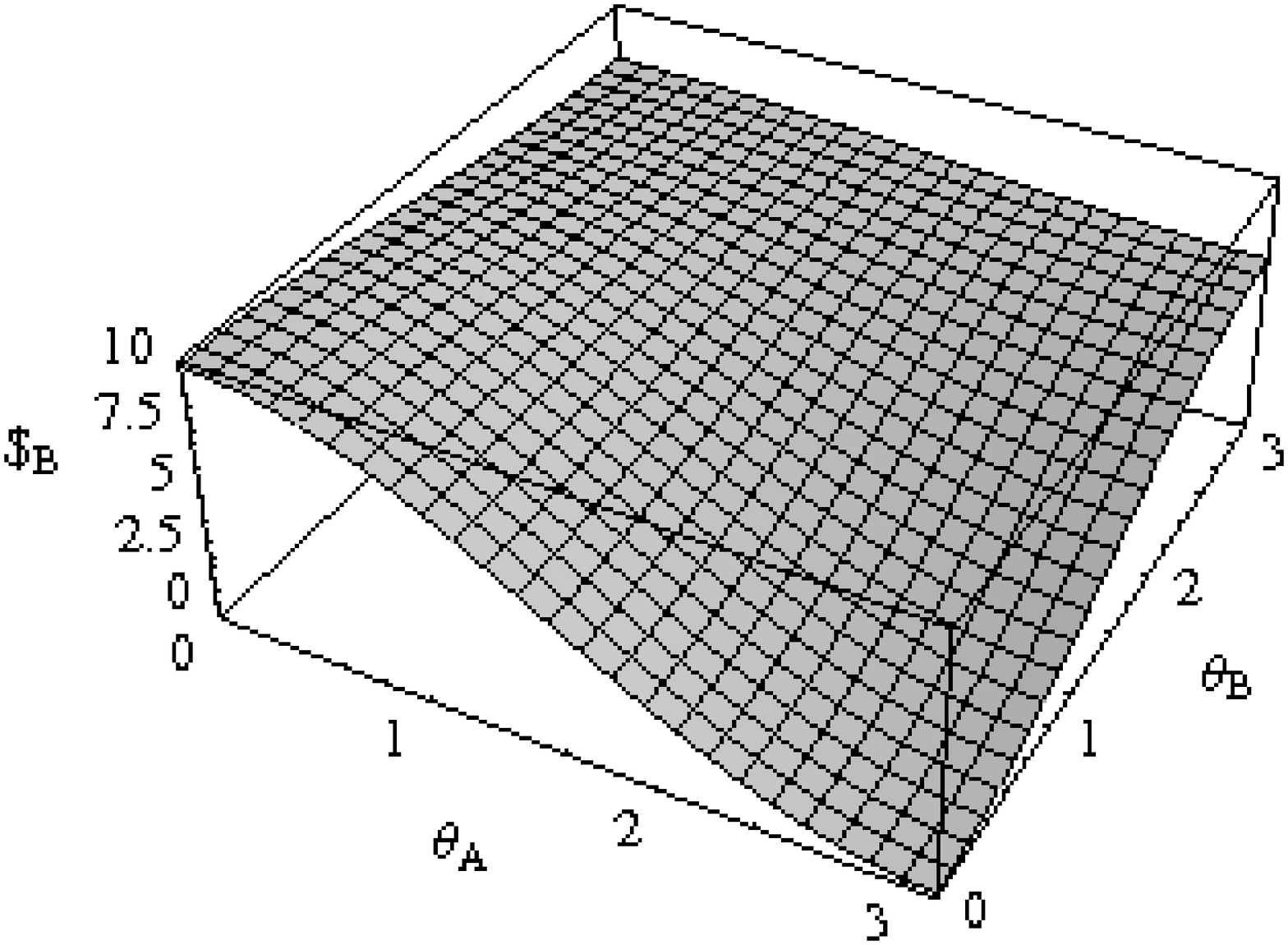}
}
\caption{SH: Payoff for Alice and Bob playing pure strategies against pure opponent strategies, classical case (1 parameter -- mixed strategies)}\label{fig:SH_klas_miesz}
\end{figure}

Another game taken into consideration is the Stag Hunt (SH). The game is described by the following payoff bimatrix:
\begin{center}
\begin{tabular}{cc}
& B \\
A & 
\begin{tabular}{c|c|c}
& C & D \\
\hline
C & (10,10) & (0,8) \\
\hline
D & (8,0) & (7,7)
\end{tabular}
\end{tabular}
\end{center}
In the classical case (fig. \ref{fig:SH_klas_miesz}) the game has two NE -- $(C,C)$ and $(D,D)$ -- one of which is Pareto optimal $(C,C)$ and both are trembling hand perfect.

\begin{figure}[x]
\centerline{
\includegraphics[width=3.0in,height=2.5in]{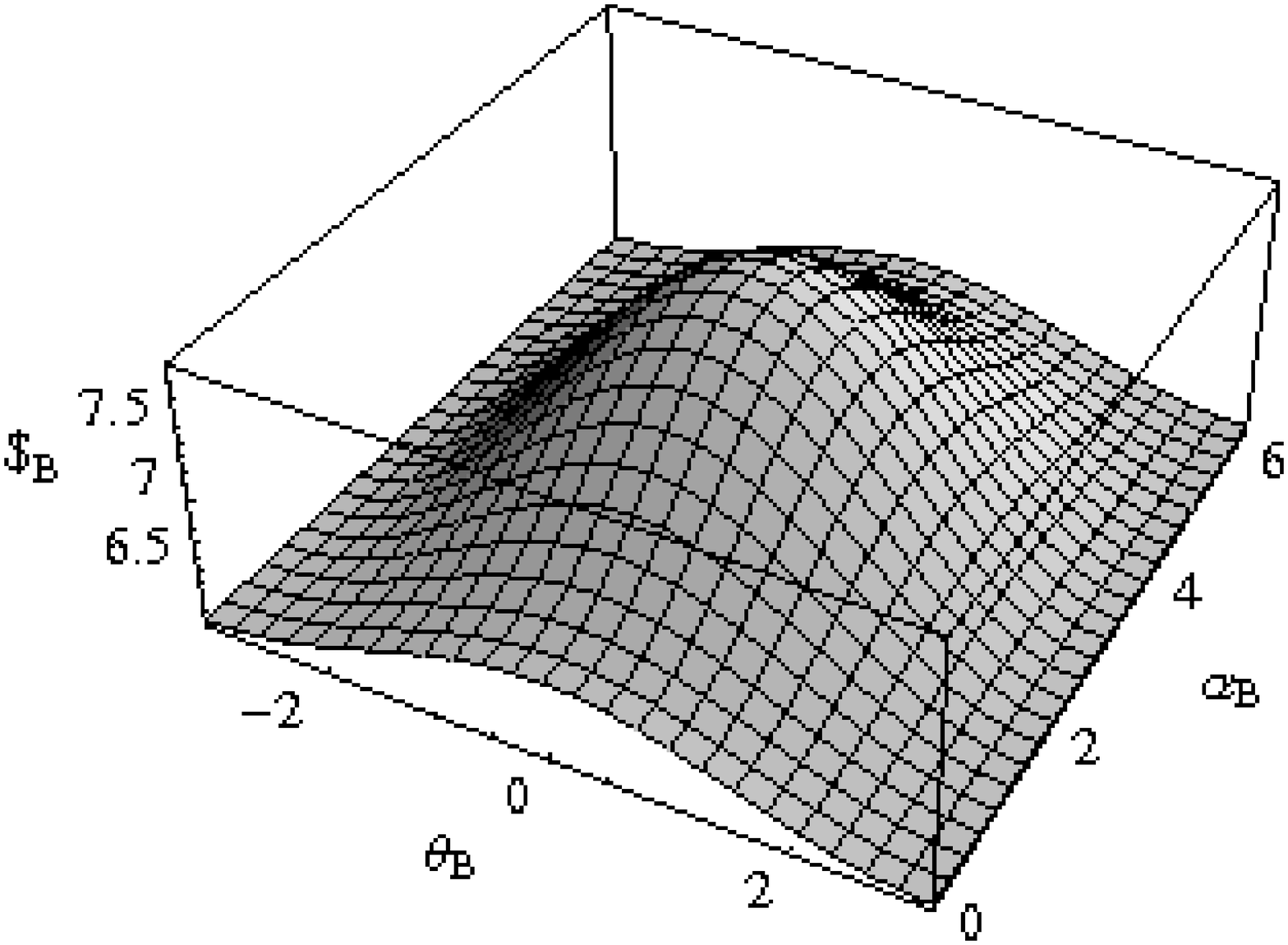}
\includegraphics[width=3.0in,height=2.5in]{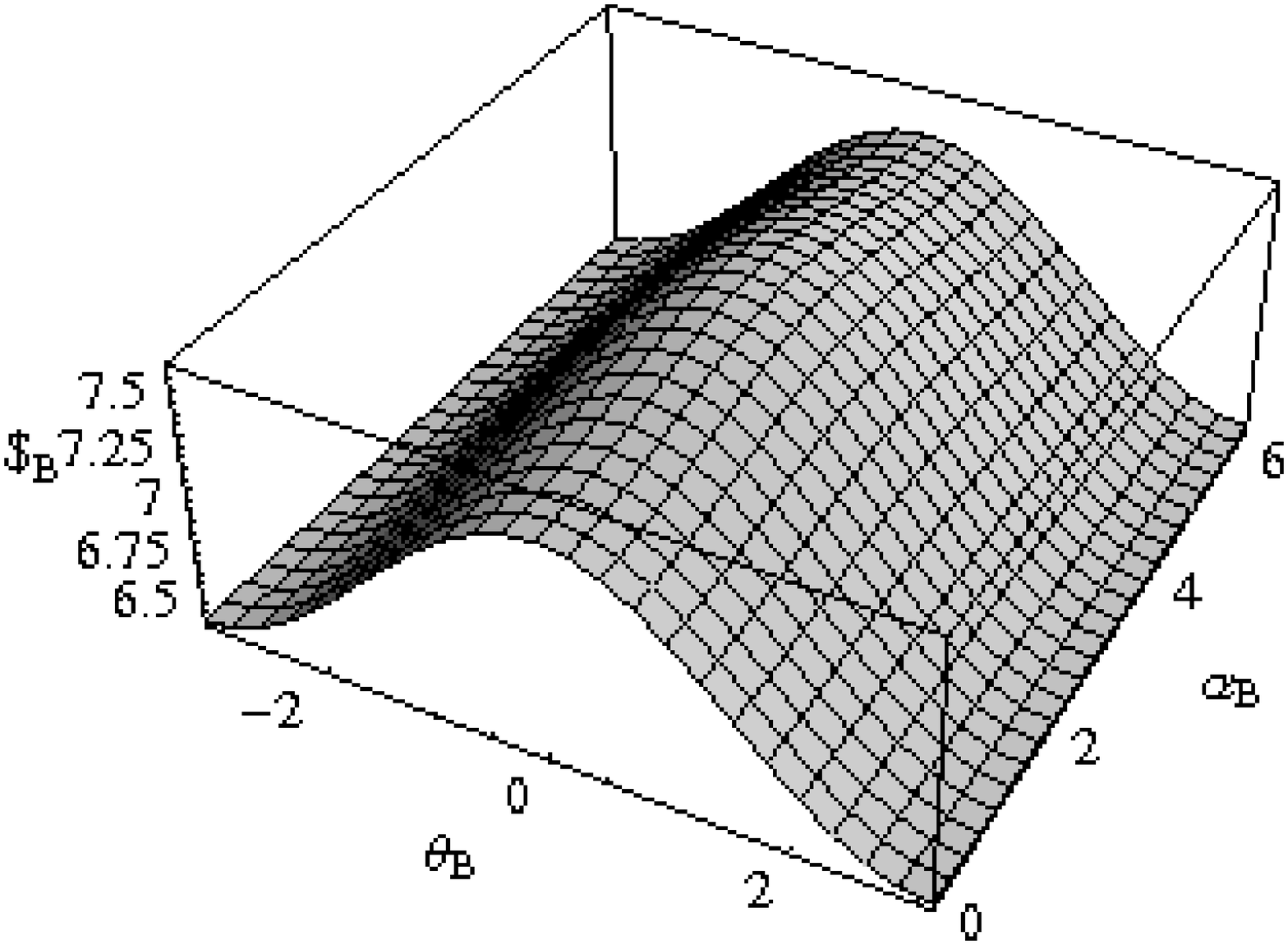}
}
\centerline{
\includegraphics[width=3.0in,height=2.5in]{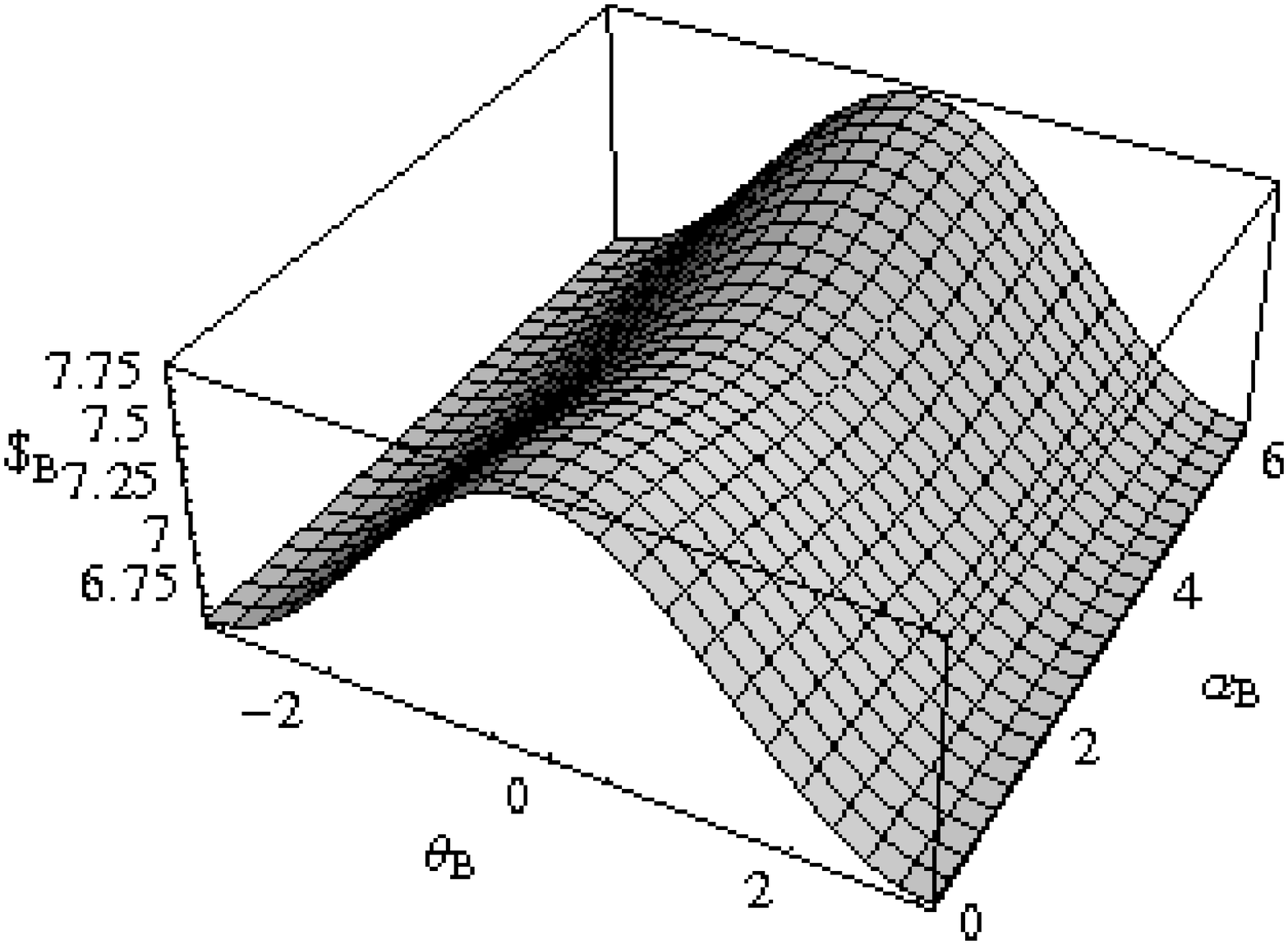}
\includegraphics[width=3.0in,height=2.5in]{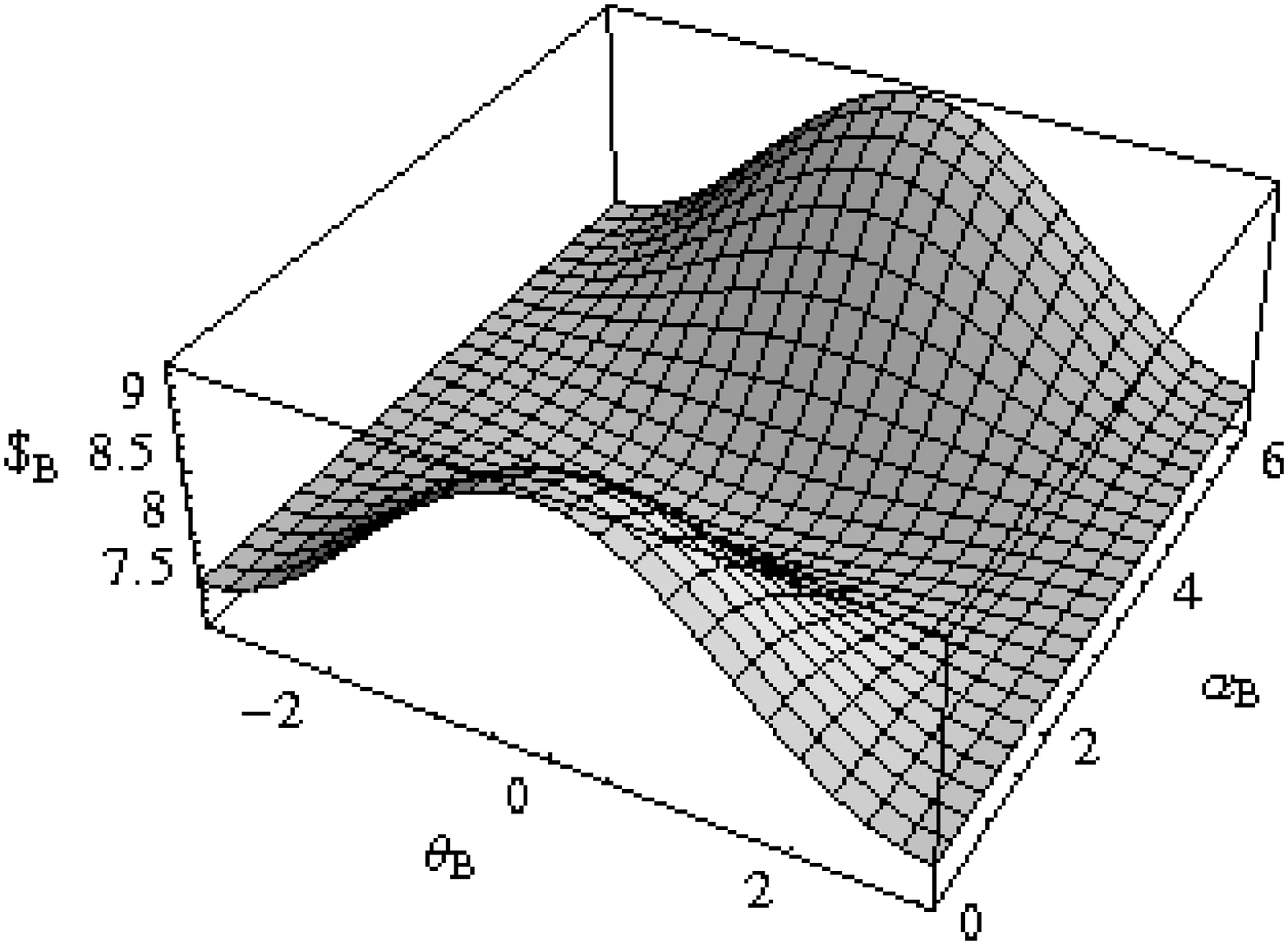}
}
\caption{SH, 2 parameters: Payoff for Bob's pure strategy against Alice's trembling C, $\kappa=1, 1.5, 1.75, 5$ -- the equilibrium appears when $\kappa>1.5$} \label{fig:SH_2par_PB_BpAtC}
\end{figure}

\begin{figure}[x]
\centerline{
\includegraphics[width=3.0in,height=2.5in]{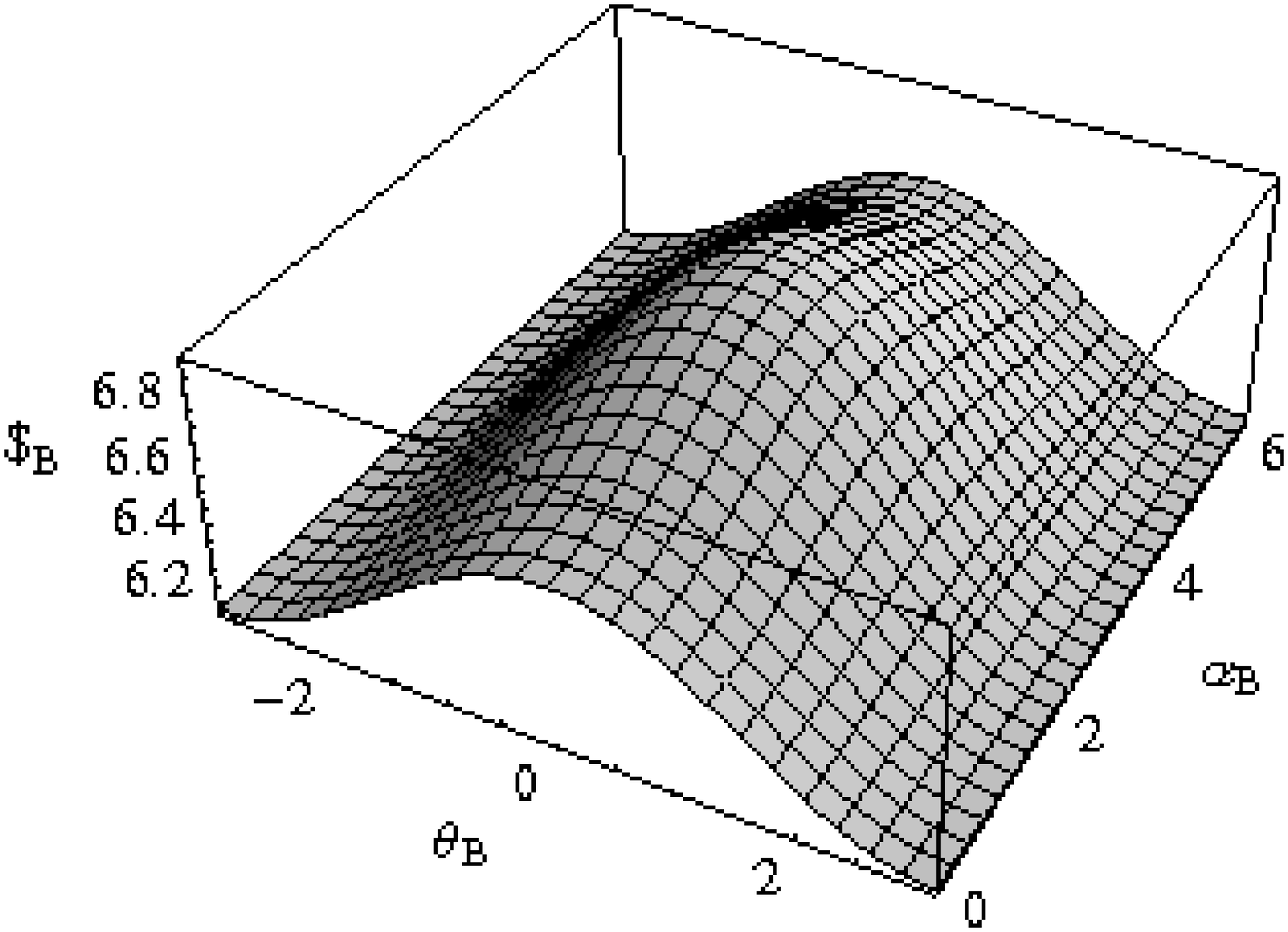}
\includegraphics[width=3.0in,height=2.5in]{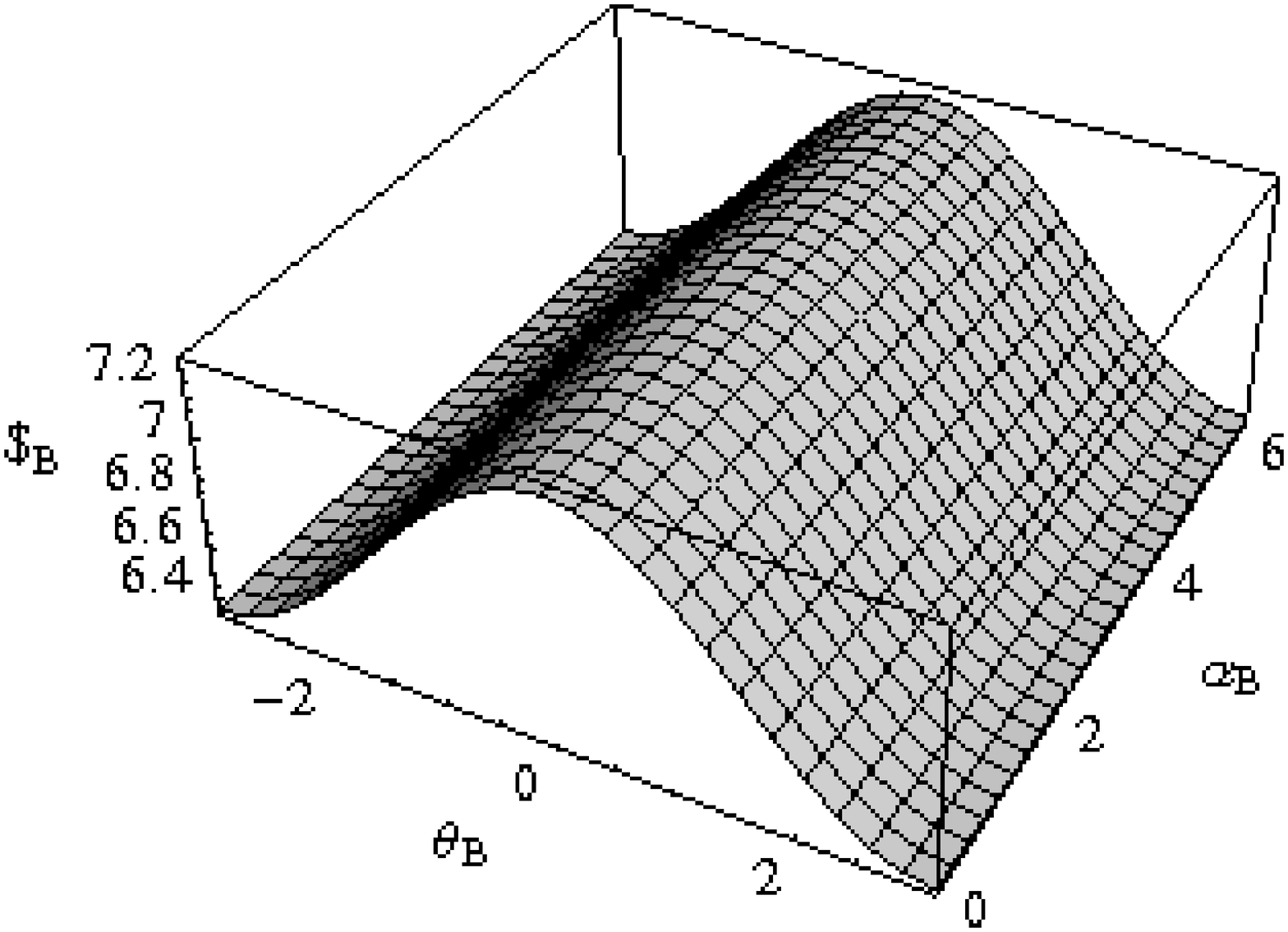}
}
\centerline{
\includegraphics[width=3.0in,height=2.5in]{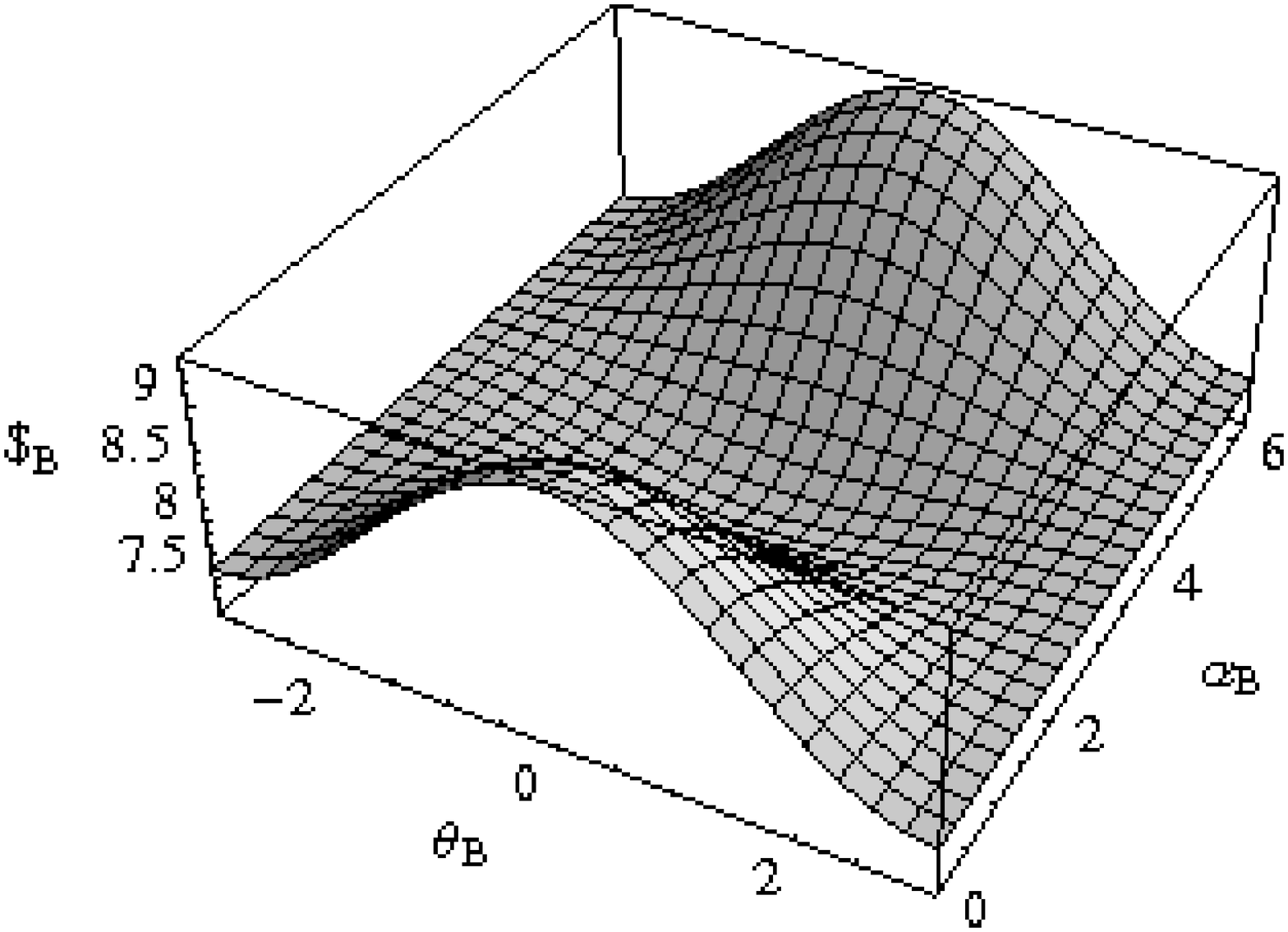}
}
\caption{SH, 3 parameters: Payoff for Bob playing pure strategies against trembling Alice's C, $\kappa=0.5, 1, 5$ -- the equilibrium appears when $\kappa>1$}\label{fig:SH_3par_PB_BpAtC}
\end{figure}

However, in the quantum case we get quite a different behaviour -- one of the equilibria disappears $(D,D)$, but a new one emerges - $(Q,Q)$. Both quantum equilibria are Pareto optimal and trembling hand perfect, but in SH there is only certain range of $\kappa$ for which the perfectness remains. When $\kappa$ is lower than threshold depending on the dimension of trembles, the stability of $(C,C)$ vanishes (figs. \ref{fig:SH_2par_PB_BpAtC}, \ref{fig:SH_3par_PB_BpAtC} -- this effect is seen for two and three parameter trembles, the best strategy response for $C$ becomes $Q$). In other words, when the errors in an implementation of this game exceed certain limit this quantum equilibrium dissapear, unlike in the case of PD and EG, where perfectness and imperfectness are present regardless of the $\kappa$ value. The perfectness of the equilibrium $(Q,Q)$ in SH does not depend on the value of $\kappa$.

\section{Conclusions}
We have extended the idea of a tremble to the quantum game theory domain and analysed three quantum versions of classical games in the context of trembling hand perfectness of their equilibria. In the case of the Prisoners' Dilemma the quantum NE found by Eisert \emph{et al.} appears to be stable with respect to trembles in both two and three parameter case. The Example Game has a weak equilibrium which is trembling hand imperfect when the game is perturbed in the two parameter space (and in the classical -- one parameter -- case). However, when the trembles in the third parameter are allowed, the weak equilibrium survives. In the Stag Hunt game one of two equilibria looses its stability when the errors exceed certain threshold and only one equilibrium remains. Due to uncertainties during the implementation of a game, three parameter trembles seem to be more accurate to predict the behaviour of equilibria, even if the strategy space is limited to two parameters. In all cases we have investigated, we find that three parameter trembles are not destroying the equilibria unless the implementation errors are small enough. However, perfectness of NE in the general case as well as other criteria of stability need to be investigated.

\section*{Acknowledgements}
This research was supported in part by the Polish Ministry of Science and Higher Education project No N519 012 31/1957

\section{Appendix A -- Quantum games}
In our calculations we use Eisert \emph{et al.} scheme \cite{eisert_thesis,eisert_et_al} of performing quantum games, with mixed states formalism. We work in the $2\times 2$ scheme, so we have two basic strategies, identity (let us describe this strategy by $C$, following the Prisoners' Dilemma game) and bit-flip ($D$), spanning two dimensional space. We generalise a classical bit to a qubit getting $SU(2)$ as a player's strategy set. Thus $2\times 2$ games are described by two qubits and unitary operations on them (in general - completely positive trace preserving maps). Players' unitary operations (quantum gates) are parametrised by three parameters in general:
\begin{equation}
U(\theta,\alpha,\beta)=\left[\begin{array}[c]{ll}
e^{i\alpha}\cos{\theta/2} & e^{i\beta}\sin{\theta/2} \\
-e^{-i\beta}\sin{\theta/2} & e^{-i\alpha}\cos{\theta/2}
\end{array}\right]
\end{equation}
Classical basic strategies are then:
\begin{eqnarray}
C=U(0,0,0)&=&\left[\begin{array}[c]{ll}
1 & 0 \\
0 & 1\end{array}\right] \\
D=U(\pi,0,0)&=&\left[\begin{array}[c]{ll}
0 & 1 \\
-1 & 0\end{array}\right]
\end{eqnarray}
Initial and final states are given by
\begin{eqnarray}
\rho_i=\pi_{CC}&=&\ket{\psi_{CC}}\bra{\psi_{CC}}\\
\rho_f&=&(A\otimes B)\rho_i (A\otimes B)^+.
\end{eqnarray}
Projectors $\pi$ are defined by the Bell states:
\begin{align}
\pi_{CC}&=\ket{\psi_{CC}}\bra{\psi_{CC}}, &\ket{\psi_{CC}}&=(\ket{00}+i\ket{11})/\sqrt{2}\\
\pi_{CD}&=\ket{\psi_{CD}}\bra{\psi_{CD}}, &\ket{\psi_{CD}}&=(\ket{01}-i\ket{10})/\sqrt{2}\\
\pi_{DC}&=\ket{\psi_{DC}}\bra{\psi_{DC}}, &\ket{\psi_{DC}}&=(\ket{10}-i\ket{01})/\sqrt{2}\\
\pi_{DD}&=\ket{\psi_{DD}}\bra{\psi_{DD}}, &\ket{\psi_{DD}}&=(\ket{11}+i\ket{00})/\sqrt{2}.
\end{align}
Payoff operators are then:
\begin{eqnarray}
P_A&=&a_{CC}\pi_{CC}+a_{CD}\pi_{CD}+a_{DC}\pi_{DC}+a_{DD}\pi_{DD}\\
P_B&=&b_{CC}\pi_{CC}+b_{CD}\pi_{CD}+b_{DC}\pi_{DC}+b_{DD}\pi_{DD}.
\end{eqnarray}
with payoff matrices
\begin{equation}
\left[\begin{array}[c]{ll}
a_{CC} & a_{CD} \\
a_{DC} & a_{DD}\end{array}\right],
\left[\begin{array}[c]{ll}
b_{CC} & b_{CD} \\
b_{DC} & b_{DD}\end{array}\right]
\end{equation}
for Alice and Bob respectively and expected payoffs given by:
\begin{eqnarray}
\avg{\$_A}&=&Tr(P_A \rho_f)\\
\avg{\$_B}&=&Tr(P_B \rho_f).
\end{eqnarray}
In the case of classical mixed strategies player $A$ can use strategy $s_i^A$ with probability $p_i^A$ and the same for player $B$ (with adequate indices changed). The average (expected) payoff for player $A$ is given by:
\begin{equation}
\bar{\$}_A=\sum_{i,j}p^A_i p^B_j \$_A(s^A_i,s^B_j)
\end{equation}
In Eisert's realisation of quantum games, classical mixed strategies are represented by operators $U(\theta,\alpha,\beta)=U(\theta,0,0)$ so $p^{A,B}_C=\cos^2 \theta_{A,B}$ and $p^{A,B}_D=\sin^2 \theta_{A,B}$.

\section{Appendix B -- von Mises-Fisher PDF}
\begin{figure}[x]
\centerline{
\includegraphics[width=3.0in,height=2.5in]{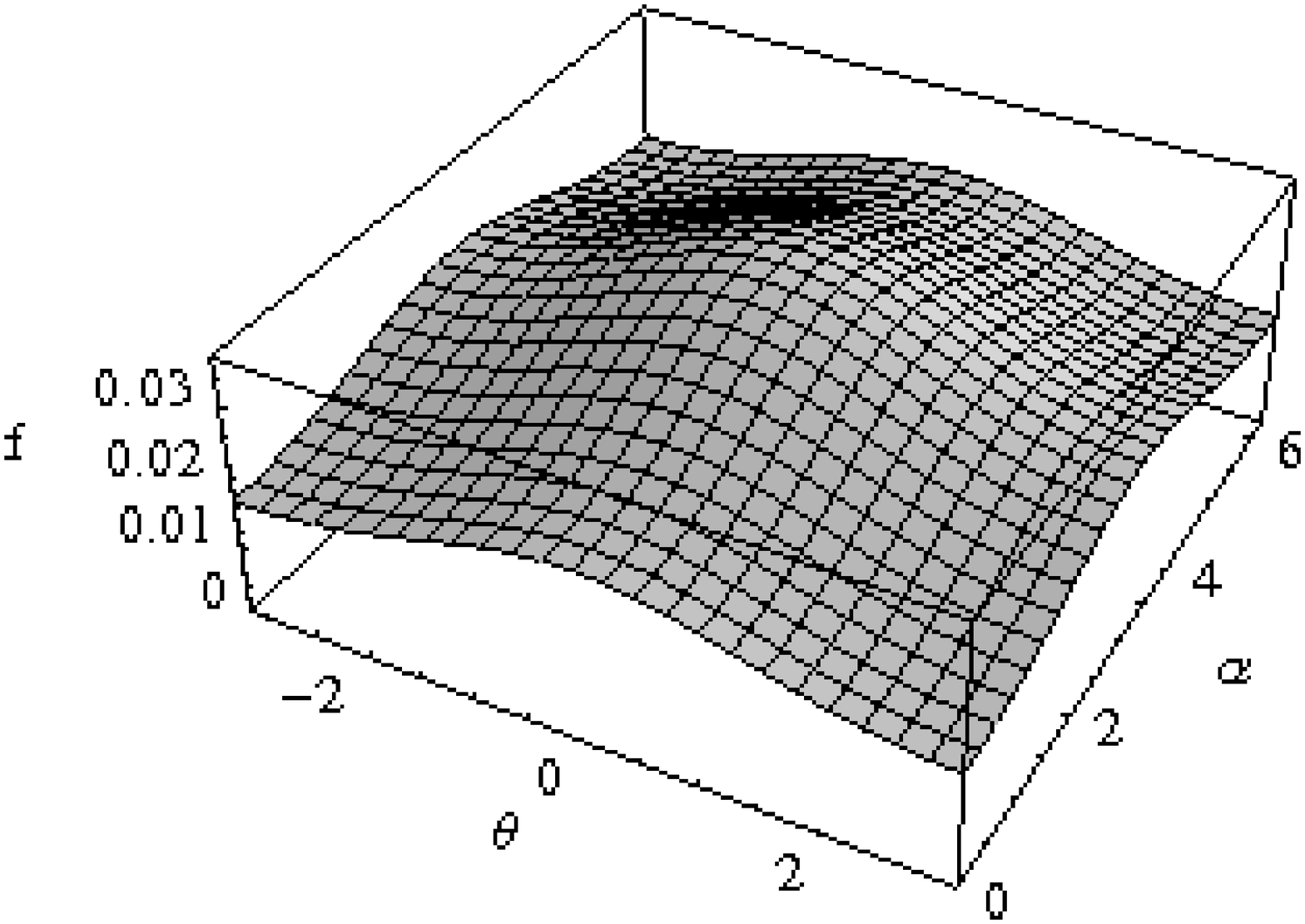}
\includegraphics[width=3.0in,height=2.5in]{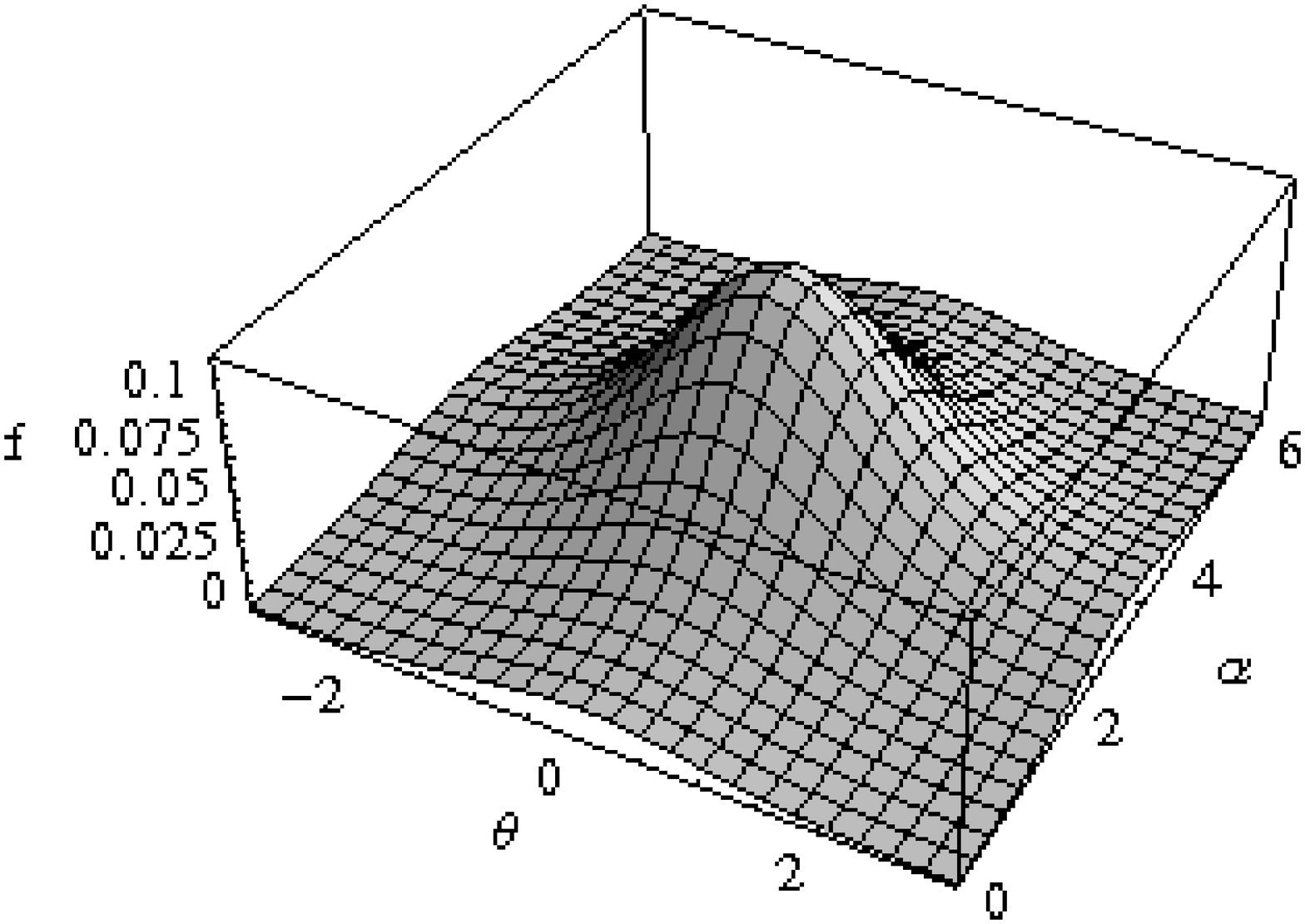}
}
\centerline{
\includegraphics[width=3.0in,height=2.5in]{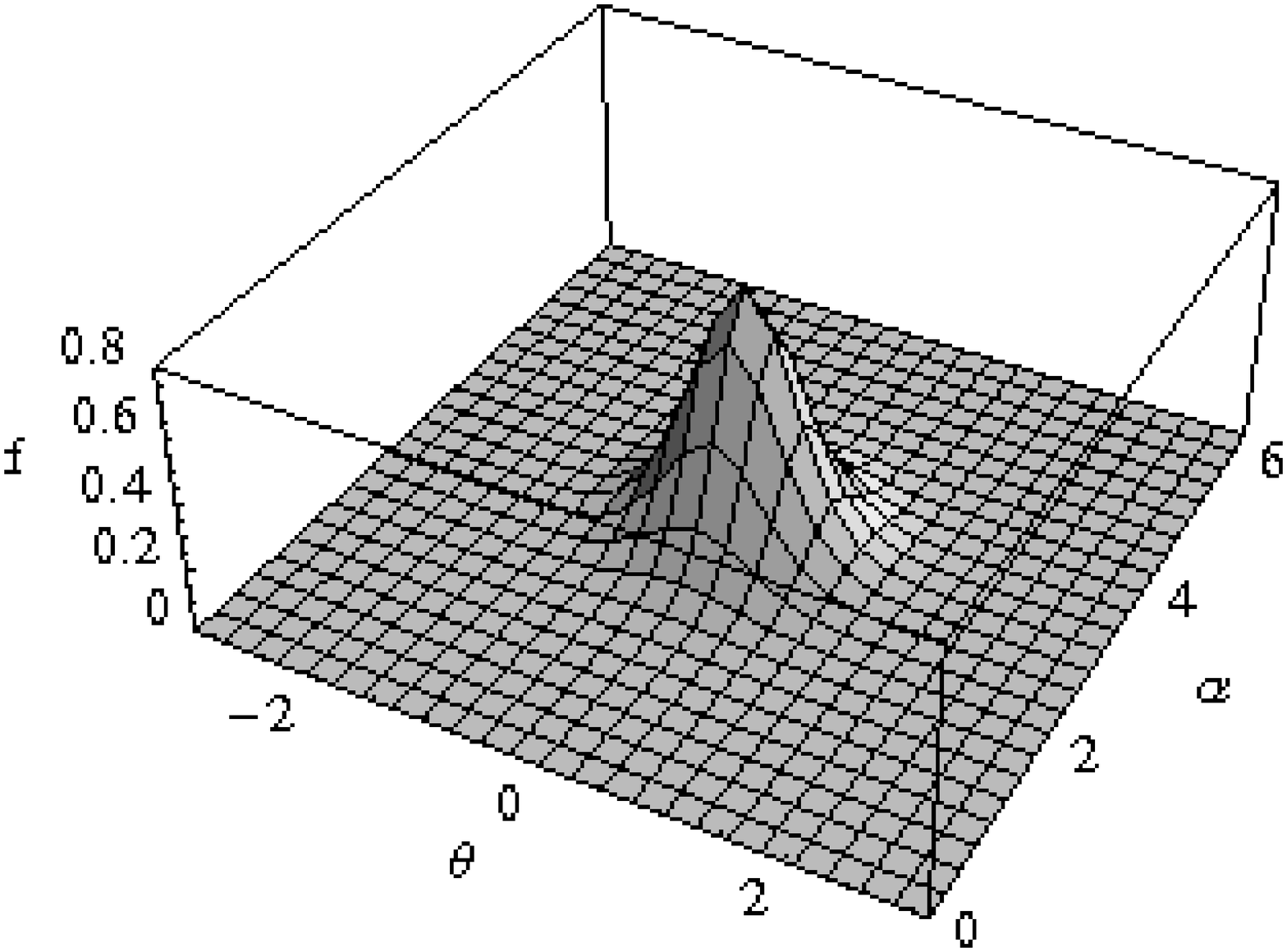}
\includegraphics[width=3.0in,height=2.5in]{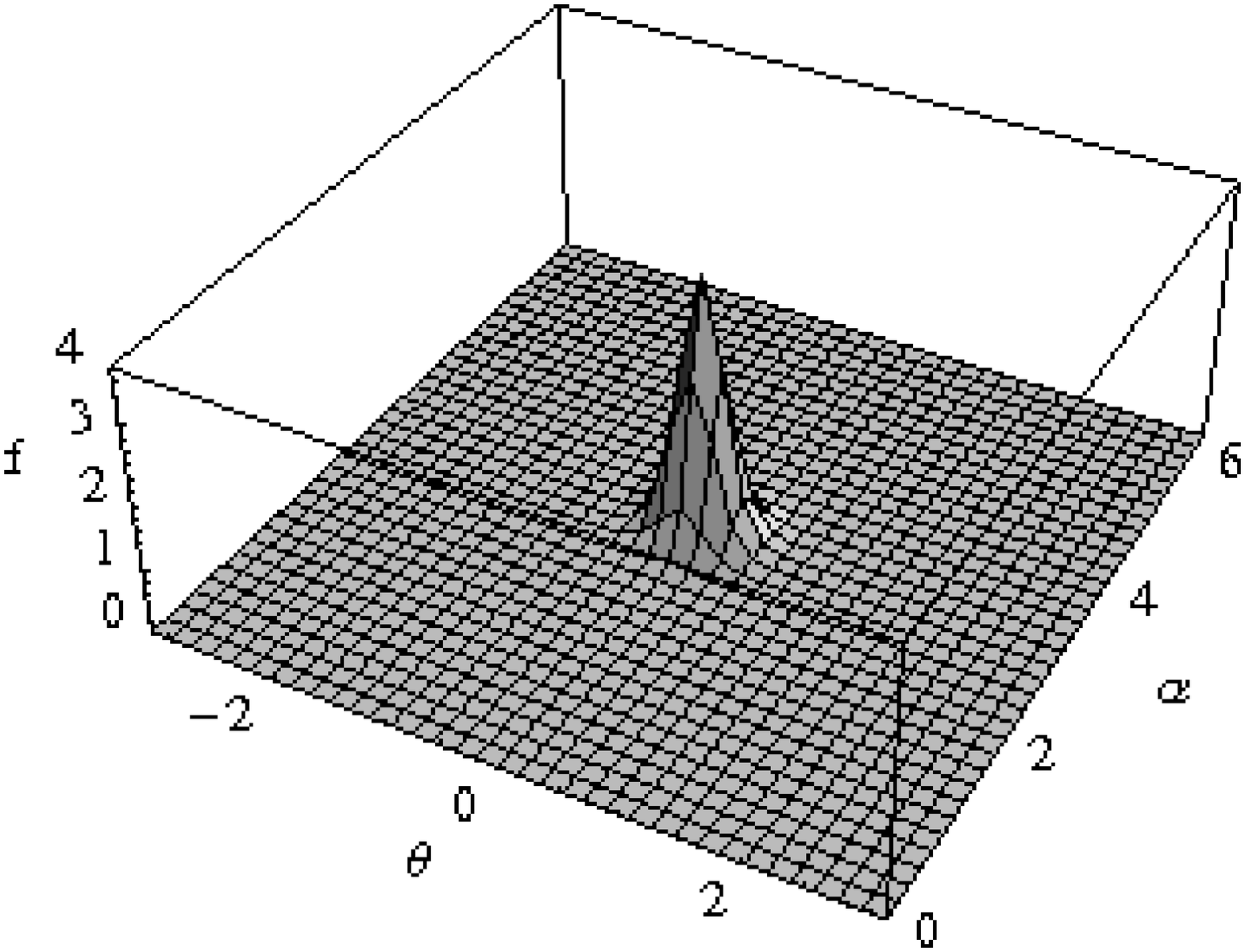}
}
\caption{2D von Mises-Fisher distribution, $\kappa=0.2,1,5,25$}\label{fig:vMF}
\end{figure}

Because our strategy spaces are spheres or tori, our PDFs representing trembles are the von Mises-Fisher distributions \cite{directional} (vMF PDFs), which are spherical versions of normal (Gauss) distribution. For a sphere $S^{p-1}$ with measure $\mu_p(\Omega)$ this distribution is of the form:
\begin{eqnarray}
f&=&c_p(\kappa)\exp(\kappa \hat{x}(\Omega)\cdot\hat{x}_0(\Omega_0))\mu_p(\Omega)\\
c_p&=&\frac{\kappa^{p/2-1}}{(2\pi)^{p/2}I_{p/2-1}(\kappa)}
\end{eqnarray}
where $c_p$ is the normalisation factor, $\kappa$ is \emph{the concentration parameter} and $I_\nu(x)$ is the modified Bessel function of order $\nu$. Versors $\hat{x}$ and $\hat{x}_0$ give a direction on a sphere, $\hat{x}_0$ is the direction to the centre of the distribution (the average value). In the case of $\kappa\to\infty$ vMF PDF goes to the Dirac delta distribution giving us the pure strategy. For $\kappa\to 0$ we acquire uniform probability distribution. Plots of von Mises distributions applied to two compact dimensions for different values of concentration parameter are given on figure \ref{fig:vMF}.

\addcontentsline{toc}{section}{References}
\bibliography{pub_THP}
\bibliographystyle{unsrt}

\begin{figure}[x]
\centerline{
\includegraphics[width=3.0in,height=2.5in]{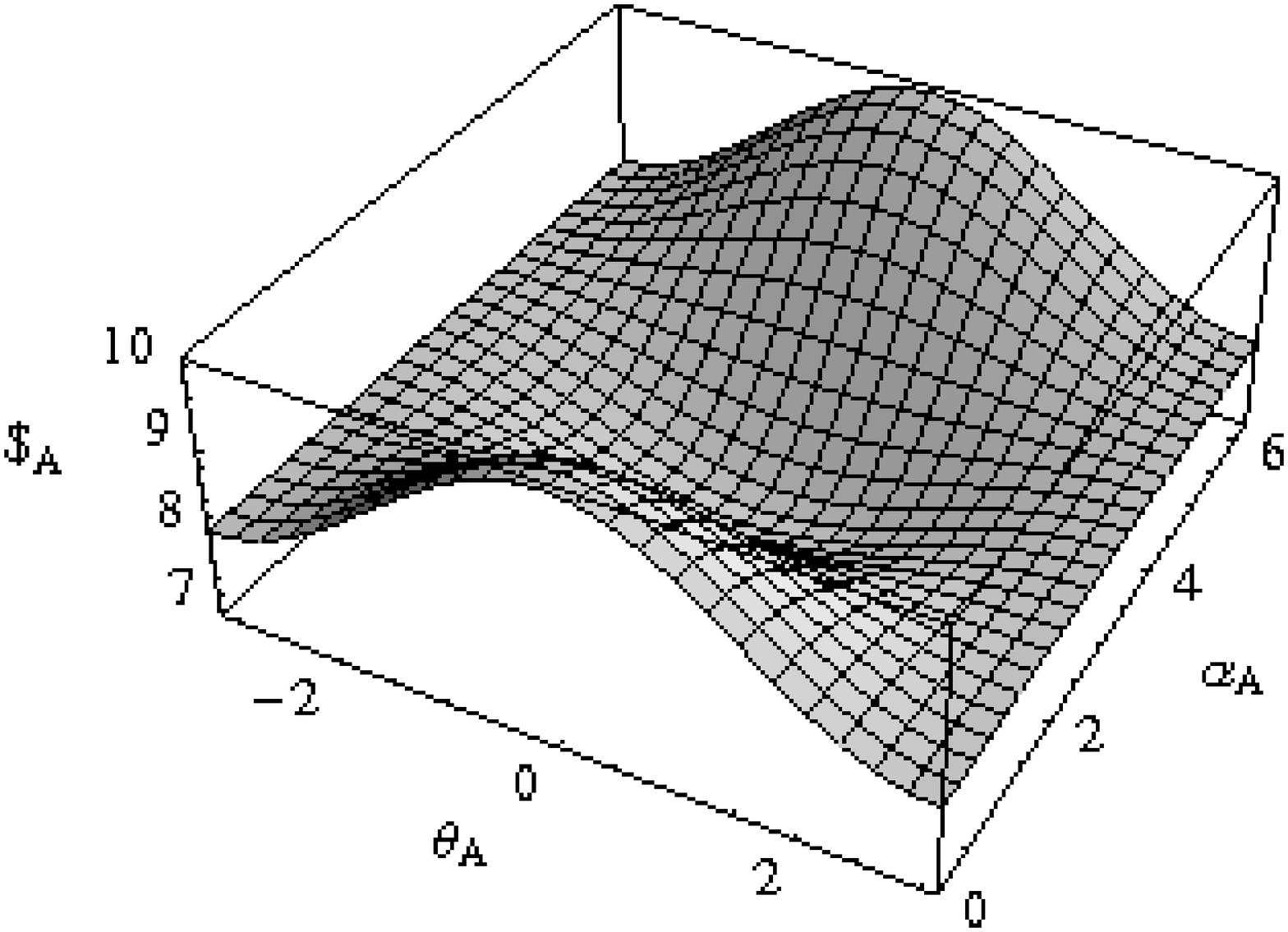}
\includegraphics[width=3.0in,height=2.5in]{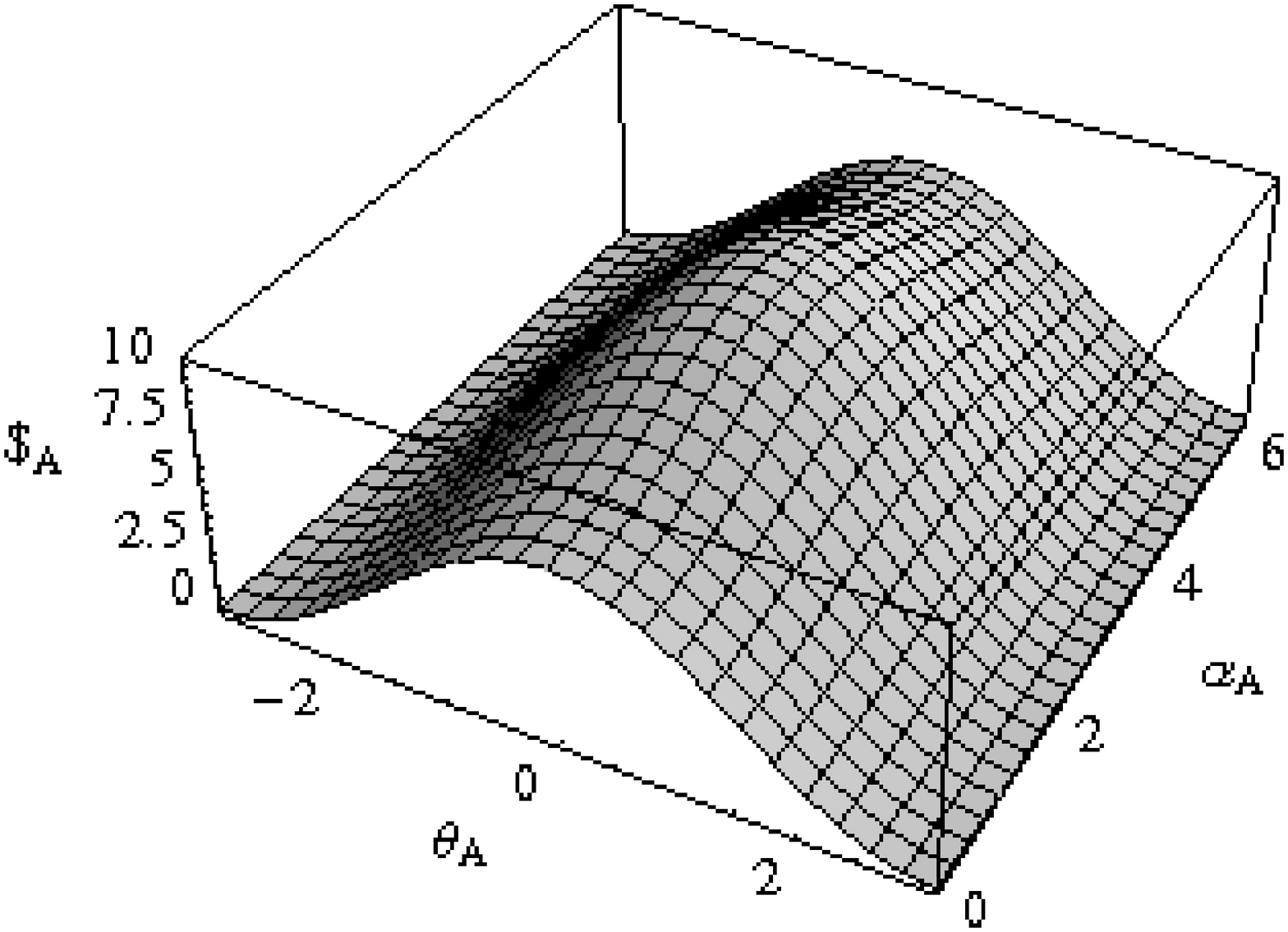}
}
\caption{SH, 2 parameters. Left: payoff for Alice playing pure strategy against Bob playing pure C. Right: payoff for Alice playing pure strategy against Bob playing pure Q (the game is symmetric)}\label{fig:SH_2par_BpCAp}
\end{figure}

\end{document}